\documentclass[prl,twocolumn,showpacs,floatfix]{revtex4}
\usepackage{times}

\usepackage{graphicx}
  \graphicspath{{./Eps/}}
\usepackage{amsmath}
\usepackage{amssymb}

\newcommand{\vc}[1]{
\textbf{\textit{#1}} }

\newcommand{\schre}{Schr\"{o}dinger }

\begin{document}

\title{Quantum and Boltzmann transport in the quasi-one-dimensional wire with rough edges}

\author{J. \surname{Feilhauer}}

\author{M. \surname{Mo\v{s}ko}}
\email{martin.mosko@savba.sk}

\affiliation{Institute of Electrical Engineering, Slovak
Academy of Sciences, 841 04 Bratislava, Slovakia}

\date{\today}

\begin{abstract}
We study electron transport in quasi-one-dimensional metallic wires. Our aim is to compare an impurity-free wire with rough edges with a smooth wire with impurity disorder. We calculate the electron transmission through the wires by
the scattering-matrix method, and we find the Landauer conductance for a large ensemble of disordered wires.
We first study the impurity-free wire whose edges have roughness with a correlation length comparable with the Fermi wave length. Simulating wires with the number of the conducting channels ($N_c$) as large as $34$ - $347$, we observe the roughness-mediated effects which are not observable for small $N_c$ ($\sim 3$ - $9$) used in previous works.
First, we observe the crossover from the quasi-ballistic transport to the diffusive one, where the ratio of the quasi-ballistic resistivity to the diffusive resistivity is $\sim N_c$ independently on the parameters of roughness.
Second, we find that transport in the diffusive regime is carried by a small effective number of open channels, equal to $\sim 6$. This number is universal - independent on $N_c$ and on the parameters of roughness.
Third, we see that the inverse mean conductance rises linearly with the wire length (a sign of the diffusive regime) up to the length twice larger than the electron localization length. We develop a theory based on the weak-scattering limit and semiclassical Boltzmann equation, and we explain the first and second  observations analytically.
For impurity disorder we find a standard diffusive behavior. Finally, we derive from the Boltzmann equation the semiclassical
electron mean-free path and we compare it
with the quantum mean-free path obtained from the Landauer conductance. They coincide for the impurity disorder, however, for the edge roughness they strongly differ, i.e., the diffusive transport in the wire with rough edges is not semiclassical. It becomes semiclassical only for roughness with large correlation length. The conductance then behaves like the conductance of the wire with impurities, also showing the conductance fluctuations of the same size.
\end{abstract}

\pacs{73.23.-b, 73.20.Fz}
\keywords{quasi one-dimensional transport, surface roughness, quantum conductance,
universal conductance fluctuations}

\maketitle


\section{I. Introduction}

A wire made of the normal metal is called mesoscopic if the wire length ($L$) is smaller than the electron coherence length \cite{Imry-book,Datta-kniha,Mello-book}. It is called quasi-one-dimensional (Q1D), if $L$ is much larger than the width ($W$) and thickness ($H$) of the wire \cite{Mello-book}. Fabrication of the Q1D wires from such metals like Au, Ag, Cu, etc., usually involves techniques like
the electron beam lithography, lift-off, and metal evaporation. These techniques always provide wires with disorder due to the grain boundaries, impurity atoms and rough wire edges \cite{Saminadayar}. Disorder scatters the conduction electrons  and limits the electron mean free path ($l$) in the wires to $\sim 10 - 100$nm \cite{Mohanty}. Of fundamental interest are the wires with $W$ and $H$ as small as $ \sim 10 - 100$nm.

 In this work the electron transport in metallic Q1D wires is studied theoretically. We compare an impurity-free wire with rough edges with a smooth wire with impurity disorder (a wire with grain boundaries will be studied elsewhere). We study the Q1D wires made of a two-dimensional (2D) conductor ($H \rightarrow 0$) of width $W$ and length $L \gg W$.
Our results are representative for wires made of a normal metal as well as of a 2D electron gas at a semiconductor heterointerface.

\begin{figure}[t]
\centerline{\includegraphics[clip,width=\columnwidth]{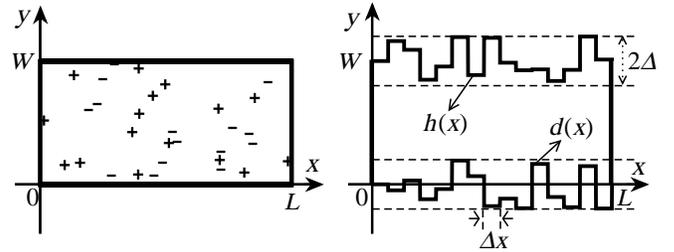}}
\vspace{-0.2cm} \caption{Wire made of the 2D conductor of width $W$ and length $L$. The figure on the left depicts the wire with impurities positioned at random with random signs of the impurity potentials. The figure on the right depicts the impurity-free wire with rough edges, where $d(x)$ and $h(x)$ are the
$y$-coordinates of the edges at $y = 0$ and $y = W$, respectively, randomly fluctuating with $x$.
The roughness amplitude is $\Delta$ and the step $\Delta x$ also means the correlation length (see the text).}
\label{Fig:1}
\end{figure}

  First we review the basic properties of the Q1D wires. Consider the electron gas confined in the two-dimensional (2D) conductor depicted in the figure \ref{Fig:1}. At zero temperature, the wave function $\varphi(x,y)$ of the electron at the Fermi level ($E_F$) is described by the \schre equation
\begin{equation}
H\varphi(x,y)=E_F\varphi(x,y)
\label{schrodgen}
\end{equation}
with Hamiltonian
\begin{equation}
 H = - \frac{\hbar^2}{2m} \left( \frac{\partial^2}{\partial x^2} + \frac{\partial^2}{\partial y^2} \right) + V \left( x,y \right)
+ U_I \left( x,y \right) ,
\end{equation}
where $m$ is the electron effective mass, $U_I(x,y)$ is the potential due to the impurities, and $V(x,y)$ is the confining potential due to the edges. Following the figure \ref{Fig:1}, the confining potential in a wire with smooth edges can be written as
\begin{eqnarray}
V(y) = \left\{
                          \begin{array}{ll}
                            0, & 0<y<W \\
                            \infty, & \mbox{elsewhere}
                          \end{array} \right.
,
\label{smoothpot}
\end{eqnarray}
while in a wire with rough edges it has to be modified as
\begin{eqnarray}
V(x,y) = \left\{
                          \begin{array}{ll}
                            0, & d(x)<y<h(x) \\
                            \infty, & \mbox{elsewhere}
                          \end{array} \right.
,
\label{skokpot}
\end{eqnarray}
 where $d(x)$ and $h(x)$ are the $y$-coordinates of the edges.
The potential of the impurities, $U_I$, is usually assumed to be a white-noise potential \cite{Mello-book}. The simplest specific choice is
\begin{equation}
U_I(x,y) = \sum_{i} \gamma \delta(x - x_i) \delta(y - y_i)
\label{delta}
\end{equation}
where one sums over the random impurity positions $[x_i,y_i]$ with a random sign of the impurity strength $\gamma$
 (see Fig. \ref{Fig:1}). Similar models of disorder like in the figure \ref{Fig:1} are commonly used in the quantum transport simulations \cite{garcia,MartinSaenz2,feilhauer,cahay,tamura}

\begin{figure}[!t]
\begin{minipage}[t]{0.46\textwidth}
\begin{center}
  \includegraphics[clip,width=1.\textwidth]{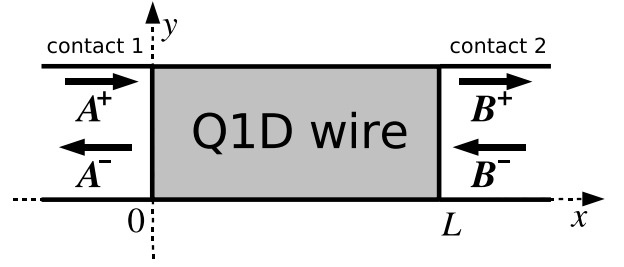}
\end{center}
\caption{The Q1D wire placed between two contacts. The bold arrows denote the wave amplitudes $\vc{A}^{+}$, $\vc{B}^{-}$ coming in the wire and the amplitudes $\vc{A}^{-}$, $\vc{B}^{+}$  coming out the wire.} \label{primka}
\end{minipage}
\end{figure}

The disordered Q1D wire is connected to two ballistic semiinfinite contacts of constant width $W$, as shown in the figure \ref{primka}. In the contacts
the electrons obey the \schre equation
\begin{equation}
 \left[ - \frac{\hbar^2}{2m} \left( \frac{\partial^2}{\partial x^2} + \frac{\partial^2}{\partial y^2} \right) + V \left( y \right)
 \right] \varphi(x,y)=E_F\varphi(x,y),
 \label{schrodka}
\end{equation}
where $V(y)$ is the confining potential given by equation \eqref{smoothpot}.
Solving equation \eqref{schrodka} one finds the independent solutions
\begin{equation} \label{subpasovafunkcia}
\varphi_n^\pm (x,y) = \\ e^{\pm ik_nx} \chi_n(y), \quad n = 1,2, \dots \infty,
\end{equation}
with the wave vectors $k_n$ given by equation
\begin{equation}
E_F = \epsilon_n + \frac{\hbar^2 k_n^2}{2m}, \quad \epsilon_n \equiv \frac{\hbar^2 \pi^2}{2mW^2} n^2,
\label{spasy}
\end{equation}
where $\epsilon_n$
is the energy of motion in the $y$-direction and
\begin{equation}
\begin{array}{c}
\chi_n(y) = \left\{
                          \begin{array}{ll}
                            \sqrt{\frac{2}{W}} \sin \left( \frac{\pi n}{W}y \right), & 0<y<W \\
                            0, & \mbox{elsewhere}
                          \end{array} \right.
\end{array}
\label{subbandfunction}
\end{equation}
is the wave function in the direction $y$.
The vectors $k_n$
in \eqref{subpasovafunkcia} are assumed to be positive, i.e.,
the waves $e^{i k_n x}$ and $e^{-i k_n x}$ describe the free motion in the positive and negative direction of the $x$-axis, respectively. The energy $\epsilon_n + \hbar^2 k_n^2/2m$ is called the $n$-th energy channel.
The channels with $\epsilon_n < E_F$ are conducting due to the real values of $k_n$ while the
channels with $\epsilon_n > E_F$ are evanescent due to the imaginary $k_n$. The conducting state $\varphi_n^+ (x,y) = e^{ik_nx} \chi_n(y)$ in the contact $1$ impinges the disordered wire from the left. It
is partly transmitted through disorder and enters the contact $2$ in the form
\begin{equation} \label{transmittedwave}
\varphi_n^+ (x,y) = \sum\limits^{\infty}_{m=1}  \
t_{mn} \ e^{ik_mx} \chi_m(y), \quad  x \geq L,
\end{equation}
where $t_{mn}(k_n)$ is the probability amplitude of transmission from $n$ to $m$. At zero temperature, the conductance of the disordered wire is
given by the Landauer formula \cite{Landauer}
\begin{equation}
G = \frac{2e^2}{h} \sum\limits^{N_c}_{n=1} T_{n}  = \frac{2e^2}{h} \sum\limits^{N_c}_{n=1} \sum\limits^{N_c}_{m=1} |t_{mn}|^2 \frac{k_m}{k_n}.
 \label{Land}
\end{equation}
where we sum over all ($N_c$) conducting channels. We note that $T_n$ is the transmission probability through disorder for the electron impinging disorder in the $n$-th conducting channel. The amplitudes $t_{mn}$ have to be calculated for specific disorder by solving the equation \eqref{schrodgen} with the asymptotic condition \eqref{transmittedwave}. In the ensemble of macroscopically identical wires disorder fluctuates from wire to wire and so does the conductance. Hence it is meaningful to evaluate \eqref{Land} for the ensemble of wires and to study the ensemble-averaged conductance $\langle G \rangle$, variance $\langle G^2 \rangle - \langle G \rangle^2$, resistance $\langle 1/G \rangle$, etc.
We now discuss a few important results of such studies. For simplicity we use the variables
$g \equiv G/(2e^2/h)$ and $\rho = 1/g$.

 First we discuss the smooth Q1D wires with disorder due to the white-noise potential $U_I$. For $L = 0$ the formula \eqref{Land}
 gives the ballistic conductance $g = N_c$. As $L$ increases, the formula \eqref{Land} first shows the classical transmission law \cite{Datta-kniha}
 \begin{equation}
\langle g \rangle = N_c \frac{\frac{\pi}{2}l}{(L+\frac{\pi}{2}l)}, \quad 0 < L \ll \xi,
 \label{Land-ballist-dif}
\end{equation}
where $N_c \gg 1$ and $\xi \simeq N_c l$ is the Q1D localization length. If $l \ll L \ll \xi$, the wire is in the diffusive regime. For $l \ll L$ and  $N_c \simeq k_F W/ \pi$ we obtain from \eqref{Land-ballist-dif} the standard expression
\begin{equation}
\langle g \rangle = \sigma_{dif} W/L, \quad \sigma_{dif} \equiv \pi n_e l/k_F,  \quad l \ll L \ll \xi,
\label{Land-dif}
\end{equation}
where $\sigma_{dif}$
is the diffusive conductivity, $k_F$ is the 2D Fermi wave vector, and $n_e = k^2_F/2\pi$ is the 2D electron density.
However, the mesoscopic diffusive conductance is also affected by weak localization. Hence, one in fact obtains from \eqref{Land} a slightly modified version of \eqref{Land-dif}, namely \cite{Mello-book,MelloStone}
\begin{equation}
\langle g \rangle = \sigma_{dif} W/L - 1/3, \quad l \ll L \ll \xi,
\label{Gweak unnorm}
\end{equation}
where the term $1/3$ is the weak localization correction typical of the Q1D wire. The mean free path $l$ in the above formulae coincides with the mean free path derived from the semiclassical Boltzmann transport equation, i.e., the quantum conductance \eqref{Land} captures the Boltzmann transport limit exactly. The mean of the two-terminal resistance, $\langle \rho \rangle \equiv \langle 1/g \rangle$, shows in absence of weak localization the diffusive behavior \cite{Datta-kniha}
\begin{equation}
\langle \rho \rangle = \rho_c + \rho_{dif}L/W, \quad 0 < L \ll \xi,
\label{Rweak unnorm}
\end{equation}
where $\rho_c = 1/N_c$ is the contact resistance and $\rho_{dif} = 1/\sigma_{dif}$ is the diffusive resistivity.
The diffusive resistance \eqref{Rweak unnorm} and diffusive conductance \eqref{Land-dif} thus coexist in a standard way: $\langle \rho \rangle \simeq 1/ \langle g \rangle$ for $\rho_{dif}\frac{L}{W} \gg \rho_c$. If we include the weak localization by means of \eqref{Gweak unnorm}, then
\begin{equation}
\langle \rho \rangle \simeq 1/\langle g \rangle \simeq \rho_{dif} \frac{L}{W} + \frac{1}{3} \rho^2_{dif} \frac{L^2}{W^2} , \quad l \ll L \ll \xi.
\label{Rweak unnorm weak}
\end{equation}
The conductance fluctuates in the diffusive regime as \cite{LeeStone,Fukuyama}
\begin{equation}
\sqrt{\mbox{var}(g)} \equiv \sqrt{\langle g^2 \rangle - \langle g \rangle^2} = \sqrt{2/15} \simeq 0.365, \quad l \ll L \ll \xi,
\label{varGa}
\end{equation}
where the numerical factor $0.365$ is typical of the Q1D wire. Finally, as $L$ exceeds $\xi$, the mesoscopic Q1D wire enters the regime of strong localization, where
$\langle g \rangle$ decreases with $L$ exponentially while $\langle \rho \rangle$ shows exponential increase \cite{Anderson,Mott}.

The formulae \eqref{Land-ballist-dif} - \eqref{varGa} hold for the wires with impurity disorder. Do they hold also for the wires with rough edges? In our present work we address this question from the first principles: we calculate the amplitudes $t_{mn}$ by
the scattering-matrix method \cite{cahay,tamura,saens} for a large ensemble of macroscopically identical disordered wires, we evaluate the Landauer conductance \eqref{Land},
and we perform ensemble averaging.

In fact, a few serious differences between the wires with rough edges and wires with impurity disorder were identified prior to our work. In the wire with impurity disorder the channels are equivalent in the sense that
$\langle T_1 \rangle = \langle T_2 \rangle \dots = \langle T_{N_c} \rangle$ \cite{Beenakker,markos}. For instance, in the diffusive regime $\langle T_n \rangle = \frac{\pi}{2}l/L$ for all channels \cite{Datta-kniha}. In the impurity-free wires with rough edges $\langle T_n \rangle$ decays fast with rasing $n$, because the scattering by the rough edges is weakest in the channel $n = 1$ and strongest in the channel $n = N_c$ \cite{Freilikher,MartinSaenz,MartinSaenz3,MartinSaenz4,Froufe}. This is easy to understand classically: in the channel $n = 1$ the electron avoids the edges by moving in parallel with them, while in the channel $n = N_c$ the motion is almost perpendicular to the edges, resulting in frequent collisions with them. As a result, $\langle T_n \rangle$ shows coexistence of the quasi-ballistic, diffusive, and strongly-localized channels \cite{SanchezGil}. Due to this coexistence, the work \cite{Freilikher} reported absence of the dependence $\langle g \rangle \propto 1/L$, suggesting that the wire with rough edges does not exhibit the diffusive conductance \eqref{Land-dif}. However, according to \cite{MartinSaenz,martinAPL}, the wire with rough edges seems to exhibit the diffusive resistance \eqref{Rweak unnorm}. In our present work these findings are examined again, but for significantly larger $N_c$ as in previous works.

First we study the impurity-free wire whose edges have a roughness correlation length comparable with the Fermi wave length.
For $L \rightarrow 0$
we observe the quasi-ballistic dependence $1/ \langle g \rangle = \langle \rho \rangle = 1/N_c + \rho_{qb} L/W$,
where $\rho_{qb}$ is the quasi-ballistic resistivity. As $L$ increases, we observe crossover to the diffusive dependence $1/ \langle g \rangle \simeq \langle \rho \rangle = 1/N^{eff}_c + \rho_{dif} L/W$, where
$\rho_{dif} \ll \rho_{qb}$ and
$1/N^{eff}_c$ is the effective contact resistance due to the $N^{eff}_c$ open channels. We find the universal results $\rho_{qb}/\rho_{dif} \simeq 0.6 N_c$  and $N^{eff}_c \simeq 6$ for $N_c \gg 1$.
 As $L$ exceeds the localization length $\xi$,
the resistance shows onset of localization while the conductance shows the diffusive dependence $1/ \langle g \rangle \simeq 1/N^{eff}_c + \rho_{dif} L/W$ up to $L \simeq 2 \xi$ and the localization for $L > 2 \xi$ only.
Finally, we find
\begin{equation}
\sqrt{\mbox{var}(g)} \equiv \sqrt{\langle g^2 \rangle - \langle g \rangle^2} \simeq 0.3, \quad l \ll L \lesssim 2\xi.
\label{varGa rough}
\end{equation}
The fluctuations \eqref{varGa rough} differ from \eqref{varGa} and were already reported in the past \cite{MartinSaenz2,Nikolic,SanchezGil,AndoTamura}.
For the smooth wires with impurities our calculations confirm the formulae \eqref{Land-ballist-dif} - \eqref{varGa}.

Moreover, we derive the wire conductivity from the semiclassical Boltzmann equation \cite{Fishman1,Fishman2,AkeraAndo}, and we compare the semiclassical
mean-free path
with the mean-free path obtained from the quantum resistivity $\rho_{dif}$. For the impurity disorder we find that the semiclassical and quantum mean-free paths coincide, which is a standard result. However, for the edge roughness the semiclassical mean-free path strongly differs from the quantum one, showing that the diffusive transport in the wire with rough edges is not semiclassical. We show that it is  semiclassical only if the roughness-correlation length is much larger than the Fermi wave length. For such edge roughness the conductance behaves like the conductance of the wire with impurities (formulae \ref{Land-ballist-dif} - \ref{varGa}), also showing the fluctuations \eqref{varGa}.

 The next section describes the scattering-matrix calculation of the amplitudes $t_{mn}$ for the impurity disorder and edge roughness. In section III, the impurity disorder and edge roughness are treated by means of the Boltzmann equation and the semiclassical Q1D conductivity expressions are derived. In section IV we show our numerical results. Moreover, the crossover from $1/ \langle g \rangle = \langle \rho \rangle = 1/N_c + \rho_{qb} L/W$ to $1/ \langle g \rangle \simeq \langle \rho \rangle = 1/N^{eff}_c + \rho_{dif} L/W$ in the wire with rough edges is derived by means of a microscopic analytical theory. The theory neglects localization but nevertheless captures the main features of our numerical results. In particular, we obtain analytically the universal results $\rho_{qb}/\rho_{dif} = \frac{\pi^3}{24}N_c$  and $N^{eff}_c \simeq 2.5$. A summary is given in section V.

\section{II. The scattering-matrix approach}

Consider the Q1D wire with contacts $1$ and $2$, shown in the figure \ref{primka}.
The wave function $\varphi(x,y)$ in the contacts
can be expanded in the basis of the eigenstates \eqref{subpasovafunkcia}. We introduce notations
$A^{\pm}_n(x) \equiv a^{\pm}_{n} e^{\pm i k_n x}$
and
$B^{\pm}_n(x) \equiv b^{\pm}_{n} e^{{\pm} i k_n x}$, where $a^\pm_n$ and $b^\pm_n$ are the amplitudes of the waves moving in the positive and negative directions of the $x$ axis, respectively.
At the boundary $x=0$
\begin{equation}
\varphi(0,y) = \sum\limits^N_{n=1} \left[ A^+_n(0) + A^-_n(0) \right] \chi_n (y),
\label{rozvojdrs1 0}
\end{equation}
while  at the boundary $x=L$
\begin{equation}
\varphi(L,y) = \sum\limits^N_{n=1} \left[ B^+_n(L) + B^-_n(L) \right] \chi_n (y),
\label{rozvojdrs2 L}
\end{equation}
where $N$ is the considered number of channels (ideally $N = \infty$).
We define the vectors $\vc{A}^\pm(0)$ and $\vc{B}^\pm(L)$ with components $A^\pm_{n=1, \dots N}(0)$ and $B^\pm_{n=1, \dots N}(L)$, respectively,
and we simplify the notations $\vc{A}^\pm(0)$ and $\vc{B}^\pm(L)$ as $\vc{A}^\pm$ and $\vc{B}^\pm$.
The amplitudes $\vc{A}^\pm$ and $\vc{B}^\pm$ are related through the matrix equation
\begin{eqnarray}
\left(
\begin{array}{c}
\vc{A}^- \\
\vc{B}^+ \\
\end{array}
\right)
=
\left[
\begin{array}{cc}
r & t' \\
t & r' \\
\end{array}
\right]
\left(
\begin{array}{c}
\vc{A}^+ \\
\vc{B}^- \\
\end{array}
\right),
\quad
S
\equiv
\left[
\begin{array}{cc}
r & t' \\
t & r' \\
\end{array}
\right]
\label{Smatrixrovnica}
\end{eqnarray}
where $S$ is the scattering matrix. Its dimensions are $2N \times 2N$ and its elements $t$, $r$, $t'$, and $r'$ are the matrices with dimensions $N \times N$. Physically, $t$ and $t'$ are the transmission amplitudes of the waves $\vc{A}^+$ and $\vc{B}^-$, respectively, while $r$ and $r'$ are the corresponding reflection amplitudes. The matrix elements of the transmission matrix $t$ are just the transmission amplitudes $t_{mn}$ which determine the conductance \eqref{Land}.

Consider two wires $1$ and $2$, described by the scattering matrices
$S_1$ and $S_2$. The matrices are defined as
\begin{equation}
S_1
\equiv
\left[
\begin{array}{cc}
r_1 & t'_1 \\
t_1 & r'_1 \\
\end{array}
\right], \quad S_2
\equiv
\left[
\begin{array}{cc}
r_2 & t'_2 \\
t_2 & r'_2 \\
\end{array}
\right].
\label{S-matica 1 and 2}
\end{equation}
Let
\begin{equation}
S_{12}
\equiv
\left[
\begin{array}{cc}
r_{12} & t'_{12} \\
t_{12} & r'_{12} \\
\end{array}
\right]
\label{S-matica12}
\end{equation}
is the scattering matrix of the wire obtained by connecting the wires $1$ and $2$ in series.
The matrix $S_{12}$ is related to the matrices $S_1$ and $S_2$ through the matrix equations \cite{Datta-kniha}
\begin{equation}
\begin{array}{l}
t_{12} = t_2[I - r'_1r_2]^{-1}t_1, \\
r_{12} = r_1 + t'_1r_2[I - r'_1r_2]^{-1}t_1, \\
t'_{12} = t'_1[I + r_2[I - r'_1r_2]^{-1}r'_1]t'_2, \\
r'_{12} = r'_2 + t_2[I - r'_1r_2]^{-1}r'_1t'_2,
\end{array}
\label{skladka}
\end{equation}
where $I$ is the unit matrix.
The equations \eqref{skladka} are usually written in the symbolic form
\begin{equation}
S_{12} = S_1\otimes S_2 .
\label{compositionlaw2}
\end{equation}

\subsection{A. Scattering matrix of smooth wire with impurity disorder}

\begin{figure}
\centering
\scalebox{0.3}{
\includegraphics*[0mm,0mm][280mm,150mm]{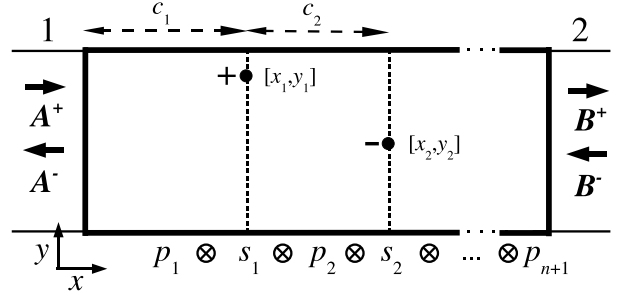}
}
\caption{Wire with the randomly positioned point-like impurities. The $n$ impurities described by the scattering matrices $s_i$ divide the wire into the $n+1$ free regions described by the matrices $p_i$. Also shown are the wave amplitudes $\vc{A}^\pm$ and $\vc{B}^\pm$.}
\label{prim}
\end{figure}

Consider the wire with impurity potential \eqref{delta}. Between any two neighboring impurities there is a region with zero impurity potential, say the region $x_{i-1}<x<x_i$,
where the electron moves along the $x$ axis like a free particle. The wire with $n$ impurities contains $n+1$ regions with free electron motion, separated by $n$ point-like regions where the scattering takes place. As illustrated in figure \ref{prim},
the scattering matrix $S$ of such wire can be obtained by applying the combination law
\begin{equation}
S = p_1\otimes s_1 \otimes p_2 \otimes s_2 \otimes\dots s_n\otimes p_{n+1},
\label{bezprek}
\end{equation}
where $p_i$ is the scattering matrix of free motion in the region $x_{i-1}<x<x_i$ and $s_i$ is the scattering matrix of the $i$-th impurity. The symbols $\otimes$ mean that the composition law \eqref{compositionlaw2} is applied in \eqref{bezprek} step by step: one first combines
the matrices $p_1$ and $s_1$, the resulting matrix is combined with $p_2$, etc.

The scattering matrix $p_i$ can be expressed as
\begin{equation}
p_i
=
\left[
\begin{array}{cc}
0 & \Phi \\
\Phi & 0 \\
\end{array}
\right],
\label{matrixoffree motion}
\end{equation}
where $0$ is the $N \times N$ matrix with zero matrix elements and $\Phi$ is the $N \times N$ matrix with matrix elements
\begin{equation}
\Phi_{mn} = e^{ik_n c_i} \delta_{mn}, \quad c_i = x_i - x_{i-1},
\label{matrixelementoffree motion}
\end{equation}
Finally, for a $\delta$-function-like impurity the scattering matrix
\begin{equation}
s_i
\equiv
\left[
\begin{array}{cc}
r & t' \\
t & r' \\
\end{array}
\right]
\label{s_i-matica}
\end{equation}
is composed of the matrices \cite{tamura,cahay}
\begin{eqnarray}
t = t' = [K + i\Gamma ]^{-1} K, \label{tt}\\
r = r' = - [K + i\Gamma]^{-1} i \Gamma,
\label{rr}
\end{eqnarray}
where $K$ and $\Gamma$ are the $N \times N$ matrices with matrix elements
\begin{equation}
K_{m n} = k_n \delta_{m n} \ \ , \ \ \Gamma_{m n} = \frac{m\gamma }{\hbar^2} \chi^{*}_{m}(y_i)\chi_n(y_i).
\label{maticky}
\end{equation}
Concerning the value of $N$, we use $N \geq N_c$ chosen in such way \cite{tamura}, that in the diffusive regime our simulation reproduces the Boltzmann-equation results.

\subsection{B. Scattering matrix of the impurity-free wire with rough edges}
\label{skokodsek}
The electrons in the impurity-free wire with rough edges are described by the \schre equation \eqref{schrodgen} with
Hamiltonian without the impurity potential, but with the confining potential $V(x,y)$ [equation \eqref{skokpot}] including the edge roughness. We specify the edge roughness as follows. We define
$x_j = j \Delta x$, where $j = 0,1,2,\dots$ and $\Delta x$ is a constant step. For $x$ between $x_{j}$ and $x_{j+1}$,
the smoothly varying functions $V(x,y)$, $h(x)$, and $d(x)$ in equation \eqref{skokpot} are replaced
by constant values $V_j(y) \equiv V(x_j,y)$, $h_j \equiv h(x_j)$, and $d_j \equiv d(x_j)$, respectively. We obtain the equation
\begin{equation}
\begin{array}{l}
V_j(y) = \left\{
                          \begin{array}{ll}
                            0, & d_j<y<h_j \\
                            \infty, & \mbox{elsewhere.}
                          \end{array} \right.
\end{array}
 \label{discreteconfinement}
\end{equation}
We assume that $d_j$ and $h_j$ vary with varying $j$ at random in the intervals $\langle -\Delta, \Delta \rangle$ and $\langle W-\Delta, W+\Delta \rangle$, respectively. This is depicted in the figure \ref{Fig:1}, where $h(x)$ and $d(x)$ fluctuate with varying $x$ by changing abruptly after each step $\Delta x$.

The wire width fluctuates with varying $x$ as well. However,
for $x$ between $x_{j}$ and $x_{j+1}$ we have the constant width
\begin{equation}
 W_j = h_j - d_j.
 \label{jdependentwidth}
\end{equation}
Consequently, the electron wave function for $x$ between $x_{j}$ and $x_{j+1}$ can be expressed in the form
\begin{equation}
\varphi^j(x,y) = \sum\limits^{N_j}_{n=1} \left[ a^+_{n} e^{i k^j_n x}  + a^-_{n} e^{- i k^j_n x} \right] \chi^j_n (y),
\label{j rozvojdrs}
\end{equation}
where $N_j$ is the considered number of channels ($N_j = \infty$ in the ideal case), the wave vectors $k^j_n$ are given by equations
\begin{equation}
E_F = \epsilon^j_n + \frac{\hbar^2 (k^j_n)^2}{2m}, \quad \epsilon^j_n \equiv \frac{\hbar^2 \pi^2}{2mW_j^2} n^2,
\label{j_spasy}
\end{equation}
\begin{equation}
\begin{array}{c}
\chi^j_n(y) = \left\{
                          \begin{array}{ll}
                            \sqrt{\frac{2}{W_j}} \sin \left[ \frac{\pi n}{W_j}(y - d_j) \right], & d_j<y<h_j \\
                            0, & \mbox{elsewhere}
                          \end{array} \right.
\end{array}
\label{j subbandfunction}
\end{equation}
are the wave functions for the $y$-direction, and the index $j$ means that the above equations hold
for $x$ between $x_{j}$ and $x_{j+1}$. The equations \eqref{j_spasy} and \eqref{j subbandfunction}
are just the equations \eqref{spasy} and \eqref{subbandfunction}, respectively, modified for the wire with rough edges.

In practice we choose $N_j$ by means of the relation \cite{saens}
\begin{equation}
N^j/N = W_j/W,
\label{j_ratio}
\end{equation}
i.e., the ratio of the channel numbers in the $j$-th region and contact regions is the same as the ratio of their widths.
We choose $N \geq N_c$ and we check, that the calculated conductance does not depend on the choice of $N$.
For $N \geq N_c$ the relation \eqref{j_ratio} ensures $N^j \geq N_c^j$, where $N_c^j$ is the number of the conducting channels in the $j$-th wire region. This makes the calculation reliable also when $N_c^j$ fluctuates with varying $j$, which happens for $\Delta$ larger than the Fermi wave length.

\begin{figure}
\centering
\scalebox{0.3}{
\includegraphics*[0mm,0mm][280mm,150mm]{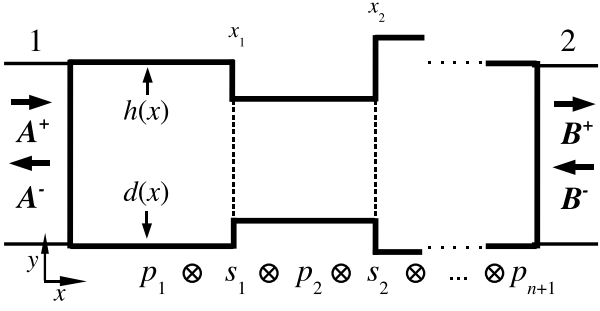}
}
\caption{The impurity-free wire with $n$ edge steps described by the scattering matrices $s_j$ and with $n+1$ free regions of constant width $W_j$, described by the scattering matrices $p_j$.}
\label{skoken}
\end{figure}

 As shown in figure \ref{skoken}, the scattering matrix $S$ of the wire with rough edges is again given by the combination law
\eqref{bezprek},
where $p_j$ is the scattering matrix of free motion in the region $x_{j-1}<x<x_j$ and $s_j$ is the scattering matrix of the $j$-th edge step.
The scattering matrix $p_j$ is given by equation \eqref{matrixoffree motion} with the matrix elements \eqref{matrixelementoffree motion} modified as
\begin{equation}
\Phi_{mn} = e^{ik^j_n \Delta x} \delta_{mn}.
\label{edgematrixelementoffree motion}
\end{equation}
Finally, we follow \cite{saens} to specify the scattering matrix $s_j$.

\begin{figure}
\centering
\includegraphics[width=8cm]{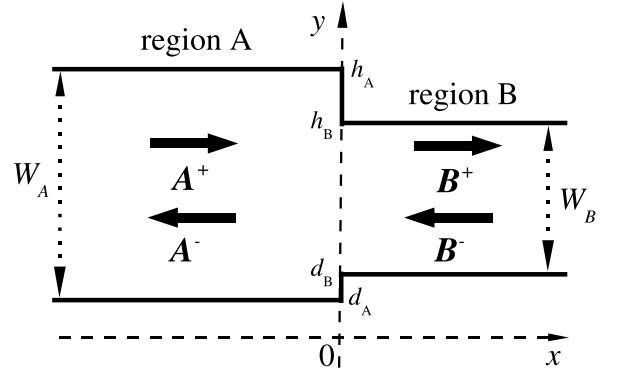}
\caption{The wire with a single edge step at $x=0$. The symbols in the figure are discussed in the text.}
\label{1skok}
\end{figure}

Consider the wire shown in figure \ref{1skok}. A single edge step at $x=0$
divides the wire into the region A ($x<0$) and region B ($x>0$). The widths of the regions A and B are $W_A = h_A - d_A$
and $W_B = h_B - d_B$, respectively. We consider the case $W_A > W_B$ and we assume, that the wire cross-section $W_A$ includes the wire cross-section $W_B$ (see the figure). In this situation, the confining potential is simply
\begin{eqnarray}
V(x,y) = \left\{
                          \begin{array}{lcr}
                            0, & x<0, & d_A <y< h_A \\
                            0, & x>0, & d_B <y< h_B \\
                            \infty, & \mbox{elsewhere.} &
                          \end{array} \right.
\end{eqnarray}
So we can use \eqref{j rozvojdrs} and write the wave function at $x = 0$ as
\begin{equation}
\varphi(0-\epsilon,y) = \sum\limits^{N_A}_{n=1} \left[ A^+_n(0) + A^-_n(0) \right] \chi^A_n (y), \quad \epsilon \rightarrow 0,
\label{edge rozvojdrs1 minusepsilon}
\end{equation}
\begin{equation}
\varphi(0+\epsilon,y) = \sum\limits^{N_B}_{n=1} \left[ B^+_n(0) + B^-_n(0) \right] \chi^B_n (y), \quad \epsilon \rightarrow 0,
\label{edge rozvojdrs2 plusepsilon}
\end{equation}
where $A^{\pm}_n(x) \equiv a^{\pm}_{n} e^{\pm i k^A_n x}$,
$B^{\pm}_n(x) \equiv b^{\pm}_{n} e^{\pm i k^B_n x}$, and $N_A$ and $N_B$ are the channel numbers in the regions A and B.
The continuity equation
$ \varphi (0-\epsilon,y) = \varphi (0+\epsilon,y)$ takes the form
\begin{eqnarray}
 \vc{A} ^+ + \vc{A}^- = C(B \rightarrow A) ( \vc{B}^+ + \vc{B}^- ),
 \label{spojkac}
\end{eqnarray}
where $C(B \rightarrow A)$ is the matrix with the matrix elements
\begin{eqnarray}
C(B \rightarrow A)_{m n} = \int_{d_A}^{h_A} \chi^{A*}_{m} (y) \chi^B_{n} (y) dy
\label{zmenab}
\end{eqnarray}
and dimensions $N_A \times N_B $. Similarly, the continuity equation
$\frac{\partial \varphi}{\partial x}(0-\epsilon,y) = \frac{\partial \varphi}{\partial x}(0+\epsilon,y)$
can be written in the form
\begin{eqnarray}
 K^A ( \vc{A}^+ - \vc{A}^- ) = C(B \rightarrow A) K^B ( \vc{B}^+ - \vc{B}^- ),
 \label{derka}
\end{eqnarray}
where $K^A$ and $K^B$ are matrices with the matrix elements
\begin{eqnarray}
 (K^A)_{mn} = k^A_n \delta_{mn}, \ \ (K^B)_{mn} = k^B_n \delta_{mn}
 \label{diagelementskakb}
\end{eqnarray}
and dimensions $N_A \times N_A $ and $N_B \times N_B$, respectively.
Combining \eqref{spojkac} and \eqref{derka} one finds the matrix equation \eqref{Smatrixrovnica}
with the scattering matrix $s_j \equiv S$ composed of the matrices
\begin{equation}
\begin{array}{l}
t = -2 M H_{BA},\\
r = -2 C(B \rightarrow A) M H_{BA} - I^A ,\\
t' = C(B \rightarrow A) [M N + I^B] ,\\
r' = M N ,\\
\end{array}
\label{trtr}
\end{equation}
where
\begin{equation}
\begin{array}{l}
H_{BA} = - (K^B)^{-1}C^T(B \rightarrow A) K^A , \\
M = [I^B - H_{BA} C(B \rightarrow A)]^{-1} ,\\
N = I^B + H_{BA} C(B \rightarrow A),
\end{array}
\label{ihmn}
\end{equation}
with $I^A$ and $I^B$ being the unit matrices and $C^T$ being the matrix obtained by transposition of the matrix $C$.
The dimensions of the matrices $t$, $r$, $t'$ and $r'$ are $N_B \times N_A $, $N_A \times N_A $, $N_A \times N_B $ and $N_B \times N_B $,
respectively.

Proceeding in a similar way one can derive $s_j$ for the situation $W_A < W_B$, with the cross-section $W_A$ included in the cross-section $W_B$. In this case $s_j$ is composed
of the matrices
\begin{equation}
\begin{array}{l}
t = C(A \rightarrow B) [M N + I^A] ,\\
r = M N,\\
t' = -2 M H_{AB},\\
r' = -2 C(A \rightarrow B) M H_{AB} - I^B ,\\
\end{array}
\end{equation}
where $M$, $H_{AB}$, and $N$ are the matrices \eqref{ihmn} with the index $A$ replaced by $B$ and vice versa.

\subsection{C. Averaging over the samples made of the building blocks}

The conductance \eqref{Land} needs to be evaluated
for a large ensemble of macroscopically identical wires because it fluctuates from wire to wire. To evaluate the ensemble-averaged results
like $\langle g \rangle$, $\langle g^2 \rangle - \langle g \rangle^2$, $\langle 1/g \rangle$, etc.,
we need to perform averaging typically over $10^3-10^4$ samples with different microscopic configurations of disorder.
Moreover, the ensemble averages are studied in dependence on the wire length. Especially for
long wires ($L \sim \xi$) with a large number of conducting channels ($N_c \sim 30 - 300$)
already the scattering-matrix calculation of a single disordered sample takes a lot of computational time.
To decrease the total computational time substantially (say a few orders of magnitude), we introduce a few efficient tricks,
partly motivated by the work \cite{Cohen}.

We recall that two scattering matrices, $S_1$ and $S_2$, can be combined by means of the operation \eqref{skladka},
written symbolically as $S_1 \otimes S_2$. This operation is associative, i.e.,
\begin{equation}
(S_1 \otimes S_2) \otimes S_3 = S_1 \otimes (S_2 \otimes S_3).
\label{associativeop}
\end{equation}
The property \eqref{associativeop} allows us to proceed as follows. We can
construct the disordered wire of length $L$ by joining a large number of short wires (building blocks), where each block has the same length ($L_b$)
and contains the same number ($n_b$) of scatterers (impurities or edge steps). The scattering matrix of such wire can be expressed as
\begin{eqnarray} \label{associativewire}
S &=& b_1 \otimes b_2 \otimes \dots  = ( p_1 \otimes s_1 \otimes \dots s_{n_b}) \otimes \nonumber\\
&{}& ( p_{n_b+1} \otimes s_{n_b+1} \otimes \dots s_{2n_b}) \otimes \dots
\end{eqnarray}
where $b_1 = p_1 \otimes s_1 \otimes \dots s_{n_b}$ is the scattering matrix of the first block,
$b_2 = p_{n_b+1} \otimes s_{n_b+1} \otimes \dots s_{2n_b}$ is the scattering matrix of the second block, etc.
In the simulation, we can create the $N_b$ different blocks (typically $N_b=100$) so that we select at random the positions of the $n_b$ scatterers in a given block. Then we evaluate the scattering matrices ($b_i$) of all $N_b$ blocks. We can now readily study
the wire lengths $L = L_b, 2L_b, 3L_b, \dots$ by applying one of the two approaches described below.

 A single sample of length $L=j L_b$ can be constructed by joining $j$ blocks, where each block is chosen at random from the $N_b$ blocks. Clearly, one can in principle construct ${(N_b)}^j$ different samples of length $j L_b$, but in practice a much smaller number of samples ($\sim 10^3 - 10^4$) is sufficient.
The $S$-matrix of each sample can be evaluated by means of \eqref{associativewire} and ensemble averaging can be performed.
Already this approach works much faster than the approach which combines in each sample the $S$-matrices of all individual scatterers.
A further significant improvement is achieved as follows.

As before, we evaluate the scattering matrices $b_i$ for all $N_b$ blocks, but we apply a more sophisticated algorithm: (i) We choose at random a single $b_i$-matrix
describing a specific sample of length $L = L_b$. (ii) Choosing at random another $b_i$ and combining it with the previous one by means of \eqref{associativewire} we obtain the $S$-matrix of a specific sample of length $L = 2 L_b$.
(iii) Choosing at random another $b_i$ and combining it with the $S$ matrix obtained in the preceding step we obtain the $S$-matrix of a specific sample of length $L = 3 L_b$. In this way we obtain a set of the Landauer conductances $\{ G(jL_b) \}_{j=1, 2, \dots}$ for a set of the specific samples
with lengths $L = L_b, 2L_b, \dots$.
Repeating the algorithm again we obtain a new set $\{ G(jL_b) \}_{j=1, 2, \dots}$. Repeating the algorithm say $10^3$ times we obtain $10^3$ various sets of $\{ G(jL_b) \}_{j=1, 2, \dots}$ and we perform ensemble averaging separately for each $j$. This approach saves a lot of time because the $S$ matrix of the sample of length $L = j L_b$ is created by combining a single $b_i$-matrix with the $S$ matrix of a sample of length $L=(j-1) L_b$.

In reality the positions of all scatterers differ from sample to sample. If we take this into account in our simulation, the ensemble-averaged results are the same (within statistical noise) as those obtained by means of our trick, but the ensemble averaging takes far much computational time. Owing to our trick we can analyze much larger systems, which is essential for a successful observation of our major results.

\section{III. Semiclassical conductivity of the Q1D wire}

In this section, the semiclassical Q1D conductivity expressions are derived both for the impurity disorder and edge roughness. Precisely, we assume that the electron motion
is semiclassical along the direction parallel with the wire and quantized in the perpendicular direction. Our approach is technically similar to the previous
 studies of the Q1D wires \cite{AkeraAndo,Palasantzas} and Q2D slabs \cite{Ando,tesanovic,Fishman1,Fishman2}.

The semiclassical conductivity of the Q1D wire is given as
\begin{equation}
 \sigma = \frac{2}{FW} \sum_n \sum_{k} \left(-\frac{e}{L}\right)\frac{\hbar k}{m} f_n(k),
 \label{sigmaINI}
\end{equation}
where $F$ is the electric field applied along the wire, $f_n(k)$ is the electron distribution function in the $n$-th conducting channel and the factor of 2 includes two spin orientations.
Similarly to the usual textbook approach, we express $f_n(k)$ as
\begin{equation}
 f_n(k) = f(E_{nk}) + \frac{eF}{\hbar}\frac{\partial f(E_{nk})}{\partial k} \tau_n(E_{nk}),
 \label{froz 1}
\end{equation}
where $f(E_{nk}) = 1/(\exp[(E_{nk} - E_F)/k_B T] + 1)$ is the equilibrium occupation number of the electron state
with energy $E_{nk} = \epsilon_n + \hbar^2 k^2/2m$ and $\tau_n(E_{nk})$ is the relaxation time.
Setting \eqref{froz 1} into \eqref{sigmaINI} we obtain at zero temperature
\begin{equation}
 \sigma = \frac{2 e^2}{\pi m W} \sum^{N_c}_{n=1} k_n \tau_n(E_F),
 \label{sigmaINI zerotemp}
\end{equation}
where $k_n$ is the Fermi wave vector in the channel $n$ [equation \ref{spasy}].
The function \eqref{froz 1}
obeys the linearized Boltzmann equation
\begin{eqnarray}
-\frac{eF}{\hbar} \frac{\partial f(E_{nk})}{\partial k} &=& \sum_{n'} \sum_{k'} W_{n,n'}(k,k') \nonumber \\
&\times& \left[ f_{n'}(k') - f_n(k) \right],
 \label{Boltz lin}
\end{eqnarray}
where
\begin{equation}
W_{n,n'}(k,k') = \frac{2 \pi}{\hbar} |\langle n' ,k' | U | n, k\rangle|^2_{\mbox{\fontsize{8}{8}\selectfont av}} \delta(E_{nk} - E_{n'k'})
\label{Wborn 1}
\end{equation}
is the Fermi-golden-rule probability of scattering from $| n, k\rangle$ to $| n', k'\rangle$, with $U$ being the scattering-perturbation potential and
$| n, k\rangle = L^{-1/2} \exp(ikx) \chi_n(y) $  being the unperturbed electron state. The
index $av$ means that $W_{n,n'}(k,k')$ is averaged over different configurations of disorder.

From \eqref{Boltz lin} we find (see the appendix A) the relaxation time
\begin{equation}
\tau_{n}(E_F) = \frac{m}{2 \pi^2 \hbar} \sum_{n'} \left( K^{-1} \right)_{nn'} k_{n'},
\label{trelax}
\end{equation}
where $K$ is the matrix with matrix elements
\begin{eqnarray}
 K_{n n'} &=&  \frac{1}{L} \sum_{k} \sum_{k'} \left[ \delta_{n n'} \sum_{\mu} |\langle n ,k | U | \mu, k'
\rangle|^2_{\mbox{\fontsize{8}{8}\selectfont av}}   \right. \nonumber \\
 \times k^2 \delta(E_{n k} &-& E_F) \delta(E_{\mu k'}-E_F) - |\langle n ,k | U | n', k' \rangle|^2_{\mbox{\fontsize{8}{8}\selectfont av}}
\nonumber\\
\times k k' \delta(E_{n k} &-& E_F) \delta(E_{n' k'} - E_F )\Bigg]
\label{KmatkaA 1}
\end{eqnarray}
The Boltzmann Q1D conductivity thus reads
\begin{equation}
 \sigma = \frac{2 e^2}{h} \frac{1}{\pi^2 W} \sum^{N_c}_{n=1} \sum^{N_c}_{n'=1} k_n k_{n'} (K)^{-1}_{n n'}.
 \label{sigmaB}
\end{equation}

\subsection{A. Semiclassical conductivity of the Q1D wire with impurities}

If we set for $U$ the impurity potential \eqref{delta}, the matrix elements \eqref{KmatkaA 1} can be expressed (see the appendix B) in the form
\begin{equation}
K_{n n'} = \frac{n_I \bar{\gamma}^2}{\pi^2 W} \left(\frac{1}{2} + \sum_{\mu =1}^{N_c} \frac{k_n}{k_{\mu}} \right) \delta_{n n'},
 \label{Kimp}
\end{equation}
where 
\begin{equation}
\bar{\gamma} = m \gamma/\hbar^2.
 \label{gammapruh}
\end{equation}
Using \eqref{Kimp} we obtain from \eqref{sigmaB} the Q1D conductivity
\begin{equation}
\sigma = \frac{2 e^2}{h} \frac{1}{\bar{\gamma}^2 n_I} \sum_{n=1}^{N_c} \frac{k^2_n}{\frac{1}{2} + \sum^{N_c}_{n'=1}
\frac{k_n}{k_{n'}}}.
 \label{sigmaIMPs}
\end{equation}
For $N_c \gg 1$ the sum in \eqref{sigmaIMPs} converges to $k_F^2 / 2$ and
\eqref{sigmaIMPs} converges to the 2D limit
\begin{equation}
\sigma = \frac{2 e^2}{h} \frac{k_F}{2} \frac{k_F}{\bar{\gamma}^2 n_I},
 \label{sigmaIMP}
\end{equation}
derivable from the 2D Boltzmann equation.

\subsection{B. Semiclassical conductivity of the Q1D wire with rough edges}

To evaluate the conductivity \eqref{sigmaB} for the wire with rough edges,
we need to determine the perturbation potential $U$
produced by the edge roughness potential in the figure \ref{Fig:1}, to set the resulting $U$ into the right hand side of \eqref{KmatkaA 1},
and to evaluate $K_{n n'}$. All this is performed in the appendix C. The result is
\begin{eqnarray}
K_{n n'} &=& \frac{\pi^2 \delta^2}{W^6} \left[ \delta_{n n'} \sum_{\mu}^{N_c} n^2 \mu^2 \frac{k_n}{k_{\mu}} \left[
\mathcal{F}(|k_{\mu}-k_n|) + \right. \right. \nonumber \\
&+& \mathcal{F}(|k_{\mu}+k_n|) ] - n^2 {n'}^2 \left[
\mathcal{F}(|k_{n}-k_{n'}|) - \right. \nonumber \\
&-& \mathcal{F}(|k_{n}+k_{n'}|)] \Bigg],
 \label{Ker}
\end{eqnarray}
where $\delta$ is the root mean square of the fluctuations of the edge coordinates $d(x)$ and $h(x)$ (see below)
and  $\mathcal{F}(q)$ is the Fourier transform of the roughness correlation function $F(x)$.

To specify $\delta$, $F(x)$, and $\mathcal{F}(q)$
we recall (see figure \ref{Fig:1}), that $d(x)$ and $h(x)$ fluctuate with varying $x$ in the intervals $\langle -\Delta, \Delta \rangle$ and $\langle W-\Delta, W+\Delta \rangle$, respectively, by changing their values abruptly after constant steps $\Delta x$.
Obviously, the values of $d(x)$ are distributed
   in the interval $\langle -\Delta, \Delta \rangle$ with the box-shaped distribution.
   Using such distribution we find, that
\begin{equation}
 \langle d(x)^2 \rangle - {\langle d(x) \rangle}^2 = \langle d(x)^2 \rangle  = \Delta^2/3 \equiv \delta^2,
 \label{rms}
\end{equation}
where the symbol $\delta$ labels the root mean square of $d(x)$. The correlation function $F(x)$ is defined as
\begin{equation}
\langle d(x_1+x) d(x_1) \rangle = \delta^2 F(x).
\end{equation}
We use the box distribution and we take into account that the step $\Delta x$ plays the role of the correlation length.
We obtain
\begin{equation}
\begin{array}{l}
F(x) = \left\{
                          \begin{array}{ll}
                            1 - |x|/\Delta x , & |x| \leq \Delta x \\
                            0, & |x| > \Delta x
                          \end{array} \right.
\end{array}
.
\end{equation}
The Fourier transform of the last equation is
\begin{equation}
 \mathcal{F}(q) = \Delta x \frac{2(1-\cos(\Delta x q))}{(\Delta x q)^2}.
 \label{Fq}
\end{equation}
The same results as for $d(x)$ hold also for $h(x)-W$, because the roughness of both edges is the same.

\begin{figure}[t]
\centerline{\includegraphics[clip,width=\columnwidth]{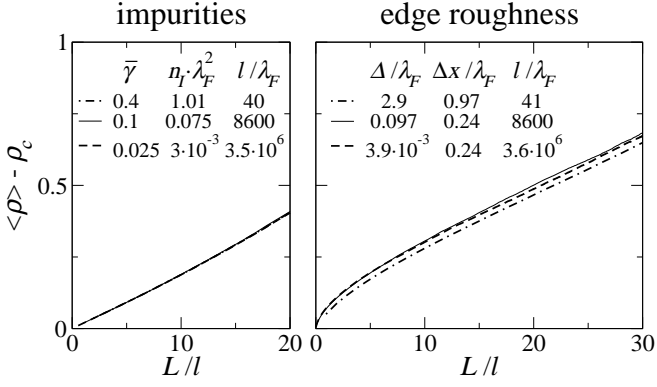}}
\vspace{-0.15cm} \caption{The mean resistance $\langle \rho \rangle$ as a function of the wire length $L$. Note that $\langle \rho \rangle$ is reduced by the contact resistance $\rho_c = 1/N_c$ and $L$ is scaled by the mean free path $l$. The left panel shows the mean resistance of the wire with impurity disorder, the right one shows the mean resistance of the wire with rough edges. The wire width $W$ is fixed
to the value $W/\lambda_F = 17.4$, which means that $N_c = 34$. The parameters of disorder are listed in the figure together with the resulting values of $l$. Determination of $l$ is demonstrated in the next figure. Note that for the Au wire ($E_F = 5.6eV$ and $m=9.109\times10^{-31}$kg) we have $\lambda_F = 0.52$nm and $W = 17.3 \lambda_F = 9$nm, with the smallest $\Delta$ being only $0.01$nm. Such small $\Delta$ is obviously not realistic, but it is used to emulate the weak roughness limit.} \label{Fig-5}
\end{figure}

\section{IV. Numerical and analytical results}

In subsection A, the quantum transport in the wires with impurity disorder and wires with rough edges
is simulated in the quasi-ballistic, diffusive, and localized regimes. In subsection B the crossover from the quasi-ballistic to diffusive regime
is explained by means of an intuitive model based on the concept of the open channels. A microscopic analytical theory of the crossover
 is given in subsection C. In subsection D, the diffusive mean-free path obtained from the Landauer conductance and mean-free path from the Boltzmann theory are compared for both types of disorder. In subsection E the wire with rough edges is studied for large roughness-correlation lengths.

In principle, all our transport results can be expressed and presented in dependence on the dimensionless variables $\Delta/\lambda_F$, $\Delta x/\lambda_F$, $W/\lambda_F$, etc., with the Fermi wave length $\lambda_F$
being the length unit. Nevertheless, in a few cases we also use the normal (not dimensionless) variables and we present the results for the material parameters $m=9.109\times10^{-31}$kg and $E_F = 5.6$eV ($\lambda_F = 0.52$nm), typical of the Au wires.

\subsection{A. Quantum transport in wires with impurities and rough edges}

We mostly simulate the wires with the number of the conducting channels being $N_c = 34$. This number emulates the limit $N_c \gg 1$ without spending too much computational time. Whenever needed we also use much larger $N_c$, the largest one being $N_c = 347$. We calculate the Landauer conductance for $10^4$ wires and
we evaluate the means.

We start with discussion of the mean resistance $\langle \rho \rangle$.
In the figure \ref{Fig-5} the mean resistance of the wire with impurity disorder is compared with the mean resistance of the wire with rough edges
for various parameters of disorder. The data obtained for various parameters tend to collapse to a single curve when plotted in dependence on the ratio $L/l$. Hence it is sufficient to discuss only the data for one specific choice of parameters.

\begin{figure}[t]
\centerline{\includegraphics[clip,width=\columnwidth]{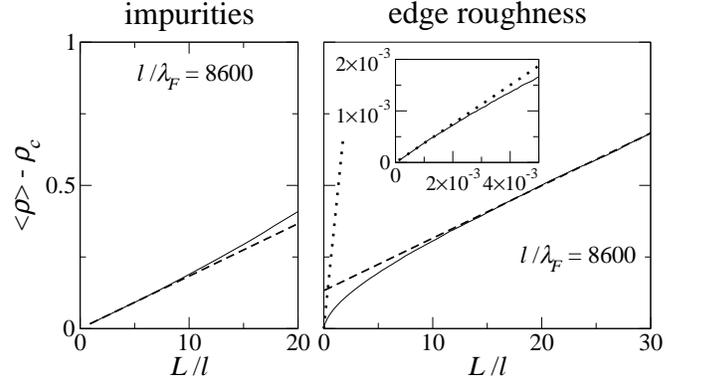}}
\vspace{-0.15cm} \caption{The full lines show the selected numerical data from the preceding figure. The dashed line in the left panel shows the linear fit $\langle \rho \rangle = \rho_c + \rho_{dif}L/W$, while the dashed line in the right panel is the linear fit $\langle \rho \rangle = \rho^{eff}_c + \rho_{dif}L/W$, with $\rho^{eff}_c$ being
the effective contact resistance. In both cases the resistivity $\rho_{dif}$ is a fitting parameter and $l$ is extracted from the Drude formula $\rho_{dif} = 2/(k_F l)$. To determine $\rho^{eff}_c$, the dashed line in the right panel is extrapolated to $L=0$. We find $\rho^{eff}_c \simeq 1/7.5$ while $\rho_c = 1/34$.
Inset shows the full line from the right panel for $L/l \ll 1$, where the transport is quasi-ballistic. The dotted lines show the linear fit $\langle \rho \rangle = \rho_c + \rho_{qb}L/W$, where the quasi-ballistic resistivity $\rho_{qb}$ is a fitting parameter. The quasi-ballistic mean free path $l_{qb}$ is given as $l_{qb} = 2/(k_F \rho_{qb})$. } \label{Fig-55}
\end{figure}

The figure \ref{Fig-55} shows the selected numerical data (full lines) from the figure \ref{Fig-5}. In accord with the textbooks \cite{Datta-kniha,Mello-book,Imry-book}, the mean resistance of the wire with impurities follows for $L \geq 0$ the standard diffusive dependence $\langle \rho \rangle = \rho_c + \rho_{dif}L/W$. However, the mean resistance of the wire with rough edges shows a more complex behavior.
For $L \rightarrow 0$ it follows the linear dependence $\langle \rho \rangle = \rho_{c} + \rho_{qb} L/W$,
where $\rho_{qb}$ is the quasi-ballistic resistivity. Only for large enough $L$ it shows crossover to the diffusive dependence $\langle \rho \rangle = \rho^{eff}_{c} + \rho_{dif} L/W$, where the resistivity $\rho_{dif}$ is much smaller that $\rho_{qb}$ and the effective contact resistance $\rho^{eff}_{c}$ strongly exceeds the fundamental contact resistance $\rho_{c}$.
In other words, the wire with rough edges shows two different linear regimes (the quasi-ballistic one and the diffusive one) separated by crossover, while the wire with impurities shows a single linear regime for the quasi-ballistic as well as diffusive transport.

The figure \ref{Fig-55} also shows that the mean resistance of the wire with impurities increases for $L/l \gtrsim 10$  slightly faster than linearly, which is due to the weak localization. However, the mean resistance of the wire with rough edges increases linearly even for $10 \lesssim L/l \lesssim 30$. The origin of this difference will become clear soon.

\begin{figure}[t!]
\begin{center}
\includegraphics[clip,width=\columnwidth]{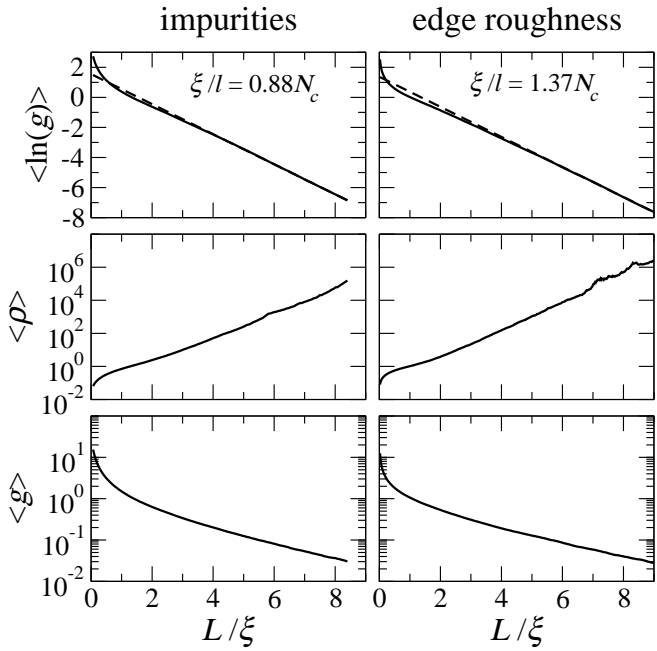}
\end{center}
\caption{Typical conductance $\langle \ln{g} \rangle$, mean resistance $\langle \rho \rangle$ and mean conductance $\langle g \rangle$ versus $L/\xi$. The results for the wire with impurity disorder (left panels) are compared with the results for
the wire with rough edges (right panels). The calculations were performed for various sets of the parameters, listed in the figure \ref{Fig-5}. The results for various parameter sets collapse to a single curve shown in a full line (see the remarks in the text). The dashed lines in the top panels show the fit $\langle \ln{g} \rangle = - L/\xi$, from which we determine the localization length $\xi$. The resulting values of $\xi$
are shown. } \label{Fig-2}
\end{figure}

The figure \ref{Fig-2} shows the numerical results (full lines) for the
the typical conductance $\langle \ln{g} \rangle$, mean resistance $\langle \rho \rangle$, and mean conductance $\langle g \rangle$ in dependence on the ratio $L/\xi$. The calculations were performed for various sets of the parameters, shown in the figure \ref{Fig-5}. The results for various sets tend to collapse to a single curve (full line) when plotted in dependence on $L/\xi$. We see for both types of disorder, that the numerical data for $\langle \ln{g} \rangle$ approach at large $L$ the dependence
$\langle \ln{g} \rangle = - L/\xi$. This is a sign of the localization \cite{LeeStoneAnderson,Fukuyama}. Fitting of the numerical data provides the values of $\xi$ shown in the figure.
  In the wire with impurities we find the result $\xi/l \simeq 0.9 N_c$, which agrees with the theoretical \cite{Thouless} prediction $\xi/l = N_c$ and with numerical studies \cite{tamura}. In the wire with rough edges we find $\xi/l \simeq 1.4 N_c$. This does not contradict the work \cite{MartinSaenz,martinAPL}, which reports $\xi/l_{1D} \simeq N_c$, but $l_{1D}$ is $\pi/2$ times larger than our $l$. Finally, due to the localization  also
$\langle \rho \rangle$ and $\langle g \rangle$ depend on $L/\xi$ exponentially at large $L/\xi$.

\begin{figure}[t!]
\centerline{\includegraphics[clip,width=\columnwidth]{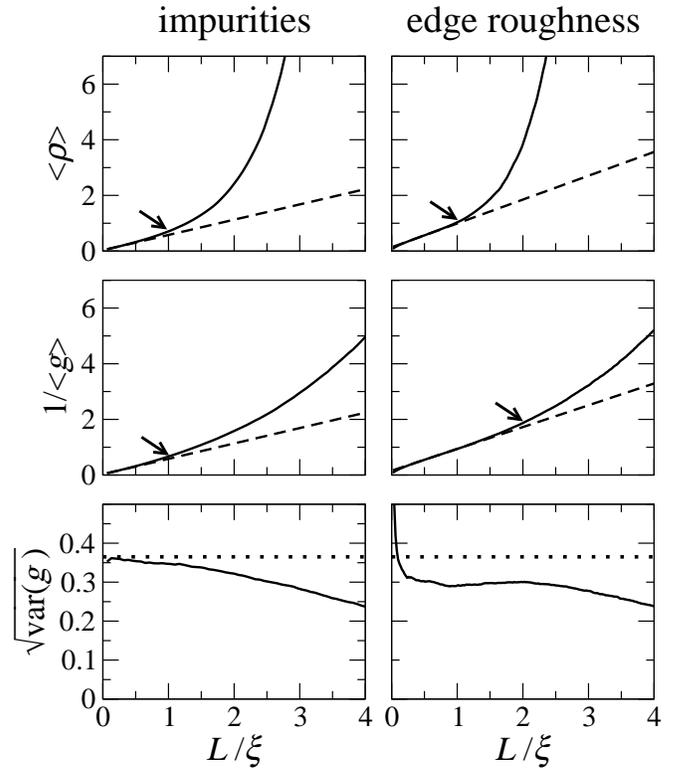}}
\vspace{-0.15cm} \caption{The full lines show our numerical data for the mean resistance $\langle \rho \rangle$, inverse mean conductance $1/ \langle g \rangle$, and conductance fluctuations $\sqrt{\mbox{var}(g)}$. The dashed lines in the top panels are the linear fits $\langle \rho \rangle = \rho_c + \rho_{dif}L/W$ (left panel) and $\langle \rho \rangle = \rho^{eff}_c + \rho_{dif}L/W$ (right panel). The dashed lines in the middle panels are the linear fits $1/ \langle g \rangle = \rho_c + \rho_{dif}L/W$ (left panel) and $1/ \langle g \rangle = \rho^{eff}_c + \rho_{dif}L/W$ (right panel). Onset of strong localization is marked by arrows at the points, where the numerical data start to deviate from the linear fit remarkably. The relative deviation from the linear fit is shown in a separate figure [figure \eqref{Fig-333}].
The dotted line in the bottom panels shows the theoretical value ($0.365$) of the conductance fluctuations, predicted
in the limit $l/\xi \ll L/\xi \ll 1$ for the white-noise disorder.
} \label{Fig-3}
\end{figure}

The figure \ref{Fig-3} show the mean resistance $\langle \rho \rangle$ and inverse mean conductance $1/ \langle g \rangle$, now in the linear scale. Concerning the mean resistance,
the wire with rough edges and wire with impurities behave similarly: $\langle \rho \rangle$ rises with $L/\xi$ linearly on the scale $l/\xi \ll L/\xi \lesssim 1$, as is typical for the diffusive regime.
However, both types of wires show a quite different $1/ \langle g \rangle$. In the wire with impurities $1/ \langle g \rangle$ rises with $L/\xi$ linearly in the interval $l/\xi \ll L/\xi \lesssim 1$, while in the wire with rough edges $1/ \langle g \rangle$ shows the linear rise with $L/\xi$ in the interval as large as $l/\xi \ll L/\xi \lesssim 2$. The $\langle g \rangle \propto 1/L$ dependence is a sign of the diffusive conductance regime, which now persists up to $L/\xi \simeq 2$ and which was not observed in \cite{Freilikher} due to the too narrow length window (as explained by the authors).

Further, the slope of the dashed lines is larger for the edge roughness than for the impurities.
 This can be understood if we write the equation $\langle \rho \rangle \simeq 1/ \langle g \rangle \simeq (2/\pi) (\xi/N_cl) (L/\xi)$ and we realize that the ratio $\xi/N_c l$ is larger for the edge roughness.

The figure \ref{Fig-3} also shows the conductance fluctuations. The fluctuations in the wire with impurities approach the universal value $0.365$, derived \cite{LeeStone,Fukuyama} in the limit $l/\xi \ll L/\xi \ll 1$ for the white-noise disorder. It is remarkable that the fluctuations in the wire with rough edges show a length-independent universal value (of size $\sim 0.3$) just in the interval $l/\xi \ll L/\xi \lesssim 2$, in which we see the linear rise of $1/ \langle g \rangle$. Coexistence of the universal conductance fluctuations with the conductance $\sim \xi/L$ is typical of the diffusive conductance regime \cite{Datta-kniha}.

In the figure \ref{Fig-3}, onset of strong localization is visible on the first glance at the points (marked by arrows), where the numerical data for $\langle \rho \rangle$  and  $1/ \langle g \rangle$ start to deviate from the linear dependence remarkably. For the edge roughness the inverse conductance shows onset of localization at $L \simeq 2 \xi$: note that the corresponding conductance fluctuations are not universal just for $L > 2 \xi$ (they decay with $L$).

\begin{figure}[t!]
\centerline{\includegraphics[clip,width=\columnwidth]{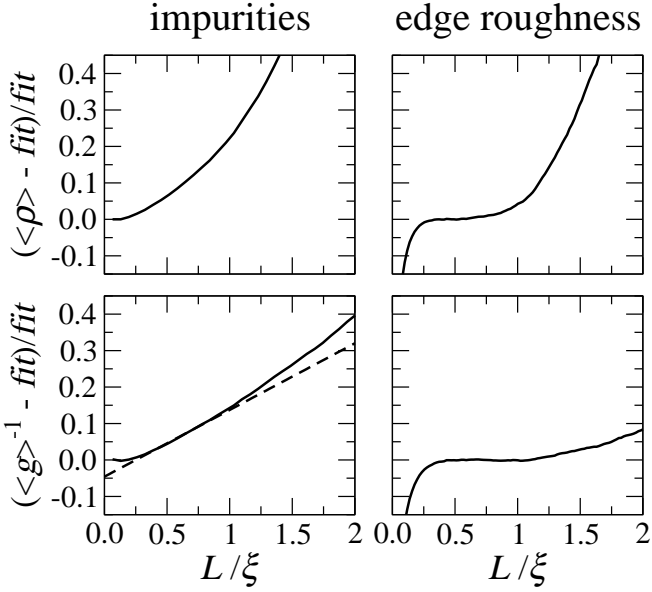}}
\vspace{-0.15cm} \caption{Numerical data for $\langle \rho \rangle$ and $1/ \langle g \rangle$ from the preceding figure, presented as a relative deviation from the linear fit. The dashed line shows the weak-localization-mediated relative deviation $\frac{1}{3} \rho_{dif} \frac{L}{W}$, shifted by replacement $L \rightarrow (L - 8l)$ to obey the limit $L \gg l$.
} \label{Fig-333}
\end{figure}
The figure \ref{Fig-333} shows the relative deviation from the linear fit, obtained from the numerical data in figure \ref{Fig-3}. Also shown is the relative deviation
$\frac{1}{3} \rho_{dif} \frac{L}{W}$, obtained from the formula \eqref{Rweak unnorm weak}. As expected, the inverse conductance of the wire with impurities exhibits for $l \ll L < \xi$ the deviation close to $\frac{1}{3} \rho_{dif} \frac{L}{W}$.
This is evidently not the case for the wire with rough edges. First, if $L \lesssim 0.2 \xi$, both $\langle \rho \rangle$ and $1/ \langle g \rangle$ show a large negative deviation due to the crossover from the quasi-ballistic to diffusive regime.
Second, if $0.2 \xi \leq L \leq  2\xi$, then $1/ \langle g \rangle$ shows the deviation as small as $\lesssim 0.08$ and almost no deviation for $0.4 \xi \lesssim L \lesssim  1.1 \xi$. In other words, $1/ \langle g \rangle$ exhibits up to $L \simeq 2 \xi$ the linear diffusive behavior with a minor nonlinear deviation. Finally, $\langle \rho \rangle$
shows a steeply increasing deviation at $L \simeq \xi$ due to the localization.

We have sofar discussed the numerical data for the wire parameters listed in figure \ref{Fig-5}. Apart from small differences due to the statistical noise, these data collapse almost precisely to the same curve, when plotted in dependence on $L/\xi$. In fact, such single-parameter scaling (dependence solely on $L/\xi$) holds exactly for weak disorder \cite{markos,Vagner,MoskoPRL}. If disorder is not weak, the data can deviate from the single-parameter scaling and the question is whether our findings hold generally.

For instance, already in the figure \ref{Fig-5} we do not see for various parameters exactly the same curves.
However, the difference between various curves is so small, that the findings extracted from one of these curves (see figure \ref{Fig-55})
are obtainable with a minor quantitative change from other curves. So we expect that the same findings hold for any (reasonable) choice of the wire parameters.
A strong support for this expectation is that the parameters used in the figure \ref{Fig-5} are very different:
the mean free paths $l$ range from $l/\lambda_F = 41$ up to $l/\lambda_F = 3.6 \times 10^6$.

\begin{figure}[t!]
\centerline{\includegraphics[clip,width=\columnwidth]{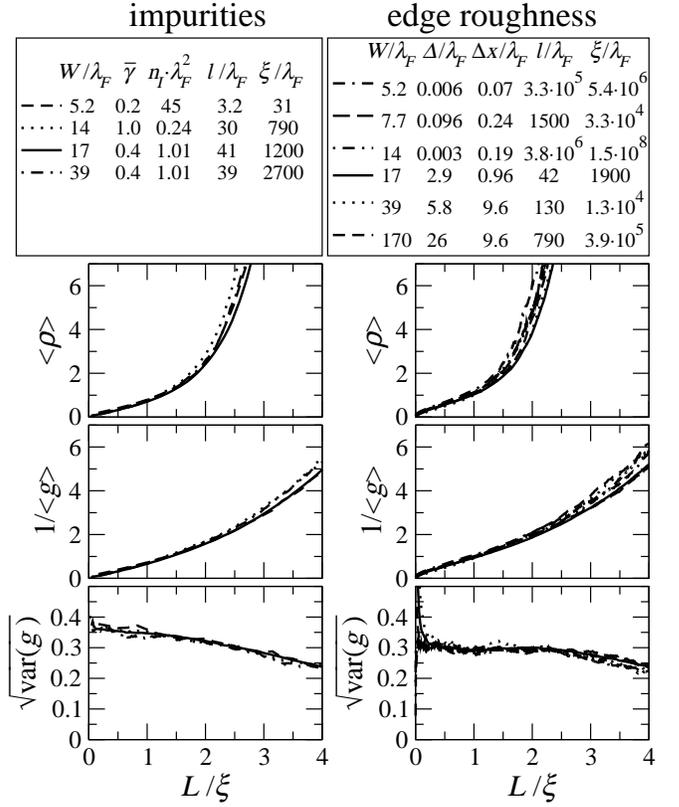}}
\vspace{-0.15cm} \caption{Mean resistance $\langle \rho \rangle$, inverse mean conductance $1/ \langle g \rangle$, and conductance fluctuations $\sqrt{\mbox{var}(g)}$ for various wire parameters.
} \label{Fig-33}
\end{figure}

The figure \ref{Fig-33} shows again the numerical data for $\langle \rho \rangle$, $1/ \langle g \rangle$, and $\sqrt{\mbox{var}(g)}$, but for various sets of the wire parameters. Obviously, the data for various sets do not collapse exactly to the same curve. This may be due to the fact that disorder is not weak, however, the resulting curves are also sensitive to how accurately we determine $\xi$. We could improve proximity of the curves in the figure \ref{Fig-33} by simulating a larger ensemble of samples and a larger wire length (in order to obtain a more accurate $\xi$). However, the presented proximity is quite sufficient in the sense that each of the curves allows to obtain the results very similar to those in figures \ref{Fig-3} and \ref{Fig-333}.
Proximity of the curves is satisfactory also with regards to the fact that the values of $\xi$ obtained for various parameters in the figure \ref{Fig-33} vary in the range of five orders of magnitude.
In this respect we can also say, that the figure \ref{Fig-33} confirms universality of the conductance fluctuations in the wire with rough edges: they are of size $\sqrt{\mbox{var}(g)} \simeq 0.3$, reported by others \cite{MartinSaenz2,Nikolic,SanchezGil,AndoTamura}.

\subsection{B. Crossover from the quasi-ballistic to diffusive transport in wires with rough edges: Intuitive analytical derivation}

         Let us examine the crossover from $\langle \rho \rangle = \rho_{c} + \rho_{qb} L/W$ to $\langle \rho \rangle = \rho^{eff}_{c} + \rho_{dif} L/W$, observed in the figure \ref{Fig-55}.
    Such crossover was not observed in the works \cite{MartinSaenz,martinAPL}, where a similar situation was studied numerically. Therefore, we first analyze the conditions of observability. We define the effective number of the open channels, $N^{eff}_c=1/\rho^{eff}_c$, and we evaluate $N^{eff}_c$ numerically (by means of the same procedure as in the figure  \ref{Fig-55}) for various wire parameters.

\begin{figure}[t!]
\centerline{\includegraphics[clip,width=\columnwidth]{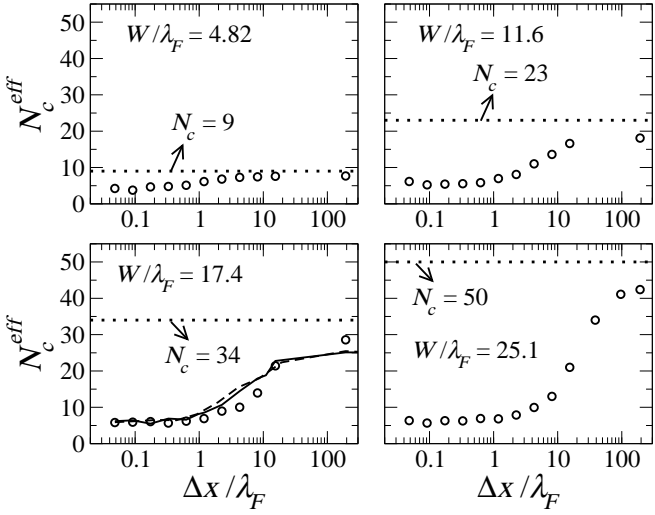}}
\vspace{-0.15cm} \caption{Effective number of open channels, $N^{eff}_c$, versus the roughness correlation length $\Delta x$ for various $N_c$ and various roughness amplitudes $\Delta$. The parameters of the edge roughness are scaled as $\Delta x/\lambda_F$ and $\Delta/W$. The results for $\Delta/W = 1/18$ are shown by open circles, the dashed and full lines show the results for $\Delta/W = 1/180$ and $\Delta/W = 1/900$, respectively.
} \label{Fig-44}
\end{figure}
   The figure \ref{Fig-44} shows $N^{eff}_c$ in dependence on the roughness-correlation length $\Delta x$ for various $N_c$ and various $\Delta$. We see that $N^{eff}_c$ reaches for $\Delta x/\lambda_F \rightarrow 0$
 a minimum value which is roughly $6$ and which depends, within our numerical accuracy, neither on $N_c$ nor on $\Delta/W$. In other words, $N^{eff}_c$ approaches for $\Delta x \rightarrow 0$ the universal value $\sim 6$.
 The calculations in \cite{MartinSaenz,martinAPL} were performed for
 $\Delta x \simeq 0.5 \lambda_F$, but only for $N_c = 5$. This value is obviously too small for noticing the existence of $N^{eff}_c \sim 6$. Hence the work \cite{MartinSaenz,martinAPL} reported the diffusive dependence $\langle 1/g \rangle \simeq 1/N_c + \rho_{dif}L/W$ rather than the dependence $\langle 1/g \rangle \simeq 1/N^{eff}_c + \rho_{dif}L/W$. Finally, for $\Delta x/\lambda_F \ggg 1$ we see that $N^{eff}_c$ approaches $N_c$. That limit is studied in the last subsection.

\begin{figure}[t!]
\centerline{\includegraphics[clip,width=\columnwidth]{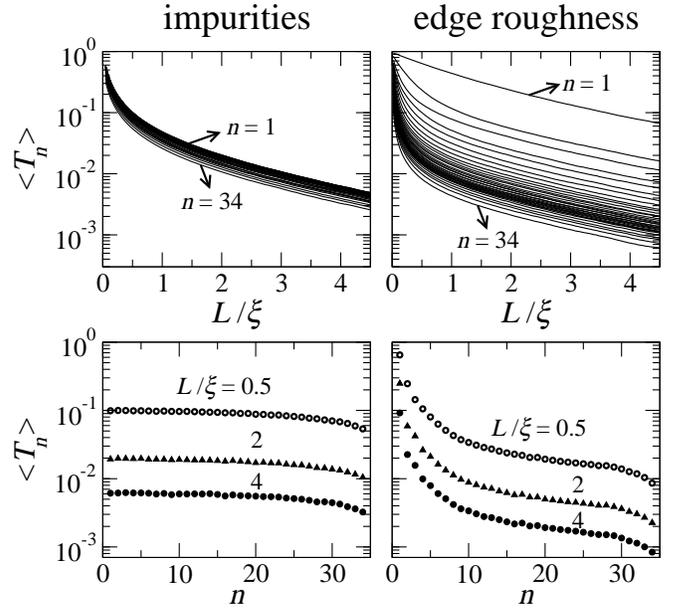}}
\vspace{-0.15cm} \caption{The top panels show the transmission probability $\langle T_n \rangle$ versus $L/\xi$ for the channel indices $n =1, 2, \dots N_c$, where $N_c = 34$. For $n$ ordered increasingly, the resulting curves are ordered
 decreasingly: the top curve shows $\langle T_{n=1} \rangle$, the bottom one shows $\langle T_{n=N_c} \rangle$. The bottom panels show $\langle T_n \rangle$ versus $n$ for various $L/\xi$.} \label{Fig-4}
\end{figure}

Further, we look at the numerical data for $\langle T_n \rangle$, shown in the figure \ref{Fig-4}. The theory based on the white-noise disorder predicts, that the conducting channels are equivalent \cite{Beenakker,markos}
in the sense that
$\langle T_1 \rangle = \langle T_2 \rangle \dots = \langle T_{N_c} \rangle$. In the figure \ref{Fig-4}, this equivalency is reasonably confirmed for the wire with impurity disorder but not for the wire with rough edges. In the wire with rough edges $\langle T_n \rangle$ decays fast with rasing $n$ which is easy to understand classically: in the channel $n = 1$ the electron avoids the edges by moving in parallel with them, while in the channel $n = N_c$ the motion is almost perpendicular to the edges, resulting in frequent collisions with them.
As a result, the $\langle T_n \rangle$ dependence in the right panels of figure \ref{Fig-4} shows for $L \simeq \xi$ the coexistence of the quasi-ballistic, diffusive, and strongly-localized channels,
 already reported in previous works \cite{Freilikher,garcia,SanchezGil}.

 Concerning the coexistence, two comments are needed. Evidently, the coexistence is not in contradiction with the fact that  all $T_n$ decay in semilogarthmic scale linearly with a single parameter $L/\xi$, when $L \gg \xi$ (see also \cite{garcia}). Further, it is clear that if
  the diffusive regime means $T_n \propto 1/L$ in all $N_c$ channels at the same value of $L$, then there is no diffusive regime but only crossover from the quasi-ballistic regime to the localization regime \cite{Freilikher}. In the preceding text we were speaking about the diffusive regime in the sense \cite{martinAPL,MartinSaenz} that $\langle \rho \rangle \propto L$ and $1/ \langle g \rangle \propto L$.

For the uncorrelated impurity disorder, the transmission $\langle T_n \rangle$ in absence of the wave interference takes the form \cite{Datta-kniha}
\begin{equation}
\langle T_n \rangle = L_n/(L_n + L),
\label{Tnkadif}
\end{equation}
where $L_n$ is the characteristic length. For the edge roughness we can adopt \eqref{Tnkadif} as an ansatz. We will prove later on that the ansatz
is indeed correct for the uncorrelated roughness. \emph{In what follows we combine the ansatz \eqref{Tnkadif} with the concept of the open channels
and we explain all major features of the crossover from the quasi-ballistic regime to the diffusive one, albeit
the formula $\eqref{Tnkadif}$ cannot capture the fact that besides the quasi-ballistic and diffusive channels there are also the localized ones.}

We asses $L_n$ from the numerical data
in figure \ref{Fig-4}. From the figure \ref{Fig-4} it is obvious, that $L_n$ in the channels with $n \gg 1$ is much smaller than $L_n$ in the channels with $n \rightarrow 1$. We emulate these findings by a simple model. We introduce the number
 $N^{eff}_c \ll N_c$ and we assume, that the channels $n=1,2,\dots,N^{eff}_c$ have the characteristic length $L_a$, while the rest of the channels has the characteristic length $L_b \ll L_a$.

 By means of the above model, we can estimate the mean resistance $\langle \rho \rangle$ and mean conductance $\langle g \rangle$ for the wire lengths $0 \leq L \lesssim \xi$. From the figure \ref{Fig-3} we see that
\begin{equation}
\langle \rho \rangle \simeq 1/ \langle g \rangle, \quad 0 \leq L \lesssim \xi.
\label{meanrho g}
\end{equation}
We therefore rely on the equation
\begin{equation}
\langle \rho \rangle \simeq \frac{1}{\langle g \rangle} = \frac{1}{\sum^{N^{eff}_c}_{n=1} \langle T_n \rangle + \sum^{N_c}_{n=N^{eff}_c+1} \langle T_n \rangle}.
\label{Rdeno}
\end{equation}
In the quasi-ballistic limit ($L \ll L_b$) the first term in the denominator of \eqref{Rdeno} is simply $N^{eff}_c$ and
the second term can be evaluated by means of \eqref{Tnkadif}. For $L_n = L_b$ we find
\begin{equation}
\langle \rho \rangle \simeq \frac{1}{N_c} + \frac{N_c - N^{eff}_c}{N^2_c L_b} L \simeq \frac{1}{N_c} + \frac{1}{N_c L_b} L,
\label{Rqblim}
\end{equation}
where the right hand side holds for $N^{eff}_c \ll N_c$.
In the diffusive regime ($l \ll L < \xi$) we evaluate the denominator of \eqref{Rdeno} by means of \eqref{Tnkadif} and we neglect the second term in the denominator assuming that $N^{eff}_c L_a \gg N_c L_b$. We get
\begin{equation}
\langle \rho \rangle \simeq \frac{L_a +L}{N^{eff}_c L_a} \simeq \frac{1}{N^{eff}_c} + \frac{1}{N^{eff}_c L_a} L.
\label{Rqblimtl}
\end{equation}
where the right hand side is the limit $L >> L_a$. \emph{This means that we assume $\langle T_n \rangle \simeq L_n/L$ for all $N^{eff}_c$ channels,
i.e., we ignore that the channel $n = 1$ is almost quasi-ballistic even at $L \simeq \xi$ (see figure \ref{Fig-4}).
Nevertheless, we succeed to obtain all major features of the crossover from the quasi-ballistic regime to the diffusive one (see mainly the next subsection).}

If we compare the formulae \eqref{Rqblim} and \eqref{Rqblimtl} with the formulae $\langle \rho \rangle = \rho_{c} + \rho_{qb} L/W$ and $\langle \rho \rangle = \rho^{eff}_{c} + \rho_{dif} L/W$, we obtain
\begin{equation}
  \rho_{qb} = \frac{W}{N_c L_b}, \quad \rho_{dif} = \frac{W}{N^{eff}_c L_a}.
\label{Rqblimtl llll}
\end{equation}
We have assumed above that $N^{eff}_c L_a \gg N_c L_b$. This means that $\rho_{qb} \gg \rho_{dif}$.
Indeed, we will see that $\rho_{qb}/ \rho_{dif} \simeq N_c$

The formula \eqref{Rqblimtl} holds only for $L \lesssim \xi$, as the equation \eqref{meanrho g} holds for $L \lesssim \xi$. However,
the formula
\begin{equation}
\frac{1}{\langle g \rangle} \simeq \frac{1}{N^{eff}_c} + \frac{1}{N^{eff}_c L_a} L
\label{GGqblimtl}
\end{equation}
 holds for $l \ll L \lesssim 2 \xi$ as we know from our numerical data. (The fact that the conductance behaves diffusively up to $L \simeq 2 \xi$ has previously been recognized from the conductance distribution \cite{MartinSaenz3}. We do not show here the conductance distributions as they are similar to those in \cite{MartinSaenz3,Froufe}.)

The formulae \eqref{Rqblim} and \eqref{Rqblimtl} show, that the crossover from $\langle \rho \rangle = \rho_{c} + \rho_{qb} L/W$ to $\langle \rho \rangle = \rho^{eff}_{c} + \rho_{dif} L/W$ is due to the channel non-equivalency. In what follows, the formulae
\eqref{Rqblim} and \eqref{Rqblimtl} will be derived from the first principles, without using the parameters $L_a$ and $L_b$.

\subsection{C. Crossover from the quasi-ballistic to diffusive transport: Microscopic analytical derivation}

We express the transmission probability $\langle T_n \rangle$ as
\begin{equation}
 \langle T_n \rangle = 1 - \sum^{N_c}_{m=1} \langle R_{mn} \rangle .
 \label{Tnondi}
\end{equation}
where
$R_{mn}$ is the probability that an electron impinging the disordered region in the $m$-th channel is reflected back into the $n$-th
channel.
In the work \cite{Freilikher} the
wire with rough edges was analyzed in the quasi-ballistic limit
and the reflection probability $ \langle R_{mn} \rangle $ was derived by means of the first order perturbation theory.
The result is
\begin{equation}
 \langle R_{mn} \rangle = 2 \times \frac{\delta^2 \kappa_m^2 \kappa_n^2}{W^2 k_m k_n} \mathcal{F}(|k_n + k_m|)L,
 \label{Rnmnondi}
\end{equation}
where $\kappa_n = (\pi/W)n$ and the factor of $2$ accounts for two edges. (In fact, the result given in \cite{Freilikher} involves a missprint. The result \eqref{Rnmnondi} can also be extracted from the backscattering length reported in \cite{Fuchs,Izrailev}.)

The mean conductance in the quasi-ballistic limit reads
\begin{equation}
\langle g \rangle = \sum_{n=1}^{N_c} \left[ 1 - \sum_{m=1}^{N_c} \langle R_{mn} \rangle \right] = N_c \left[ 1 - \frac{L}{\frac{\pi}{2}l_{qb}} \right],
 \label{Gqb}
\end{equation}
where $l_{qb}$ is the quasi-ballistic mean free path:
\begin{equation}
 l_{qb} = \frac{2}{\pi} N_c \left[ \sum_{m=1}^{N_c} \sum_{n=1}^{N_c} \frac{2 \Delta^2 \kappa_m^2 \kappa_n^2}{3W^2 k_m k_n} \mathcal{F}(|k_n + k_m|) \right]^{-1} ,
 \label{lqb}
\end{equation}
where $\delta^2 = \Delta^2/3$. Note that the quasi-ballistic limit \eqref{lqb} contains only the backscattering
contribution $\propto \mathcal{F}(|k_n + k_m|)$, while in the diffusive regime (equation \ref{Ker}) also the forward-scattering contribution $\propto \mathcal{F}(|k_n - k_m|)$ is present. The mean resistance in the quasi-ballistic limit is
\begin{equation}
\langle \rho \rangle = \frac{1}{\langle g \rangle} = \frac{1}{N_c} + \frac{1}{N_c \frac{\pi}{2}l_{qb}}L.
 \label{Rqb}
\end{equation}

\begin{figure}[t]
\begin{center}
\includegraphics[clip,width=0.9\columnwidth]{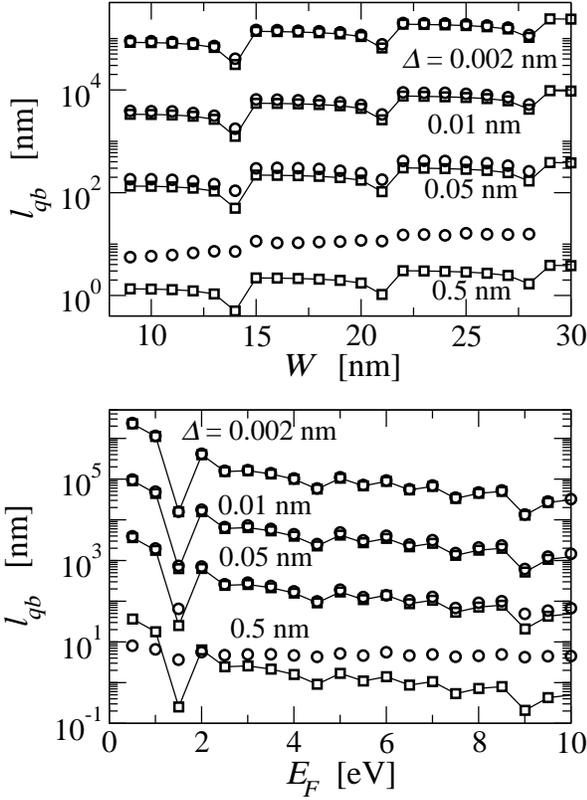}
\end{center}
\caption{Top panel: Quasi-ballistic mean free path $l_{qb}$ in the Au wire as a function of the wire width $W$ for various roughness amplitudes $\Delta$. The roughness correlation length is fixed to $\Delta x = 0.125nm$, the Au material parameters are $m=9.109\times10^{-31}$kg and $E_F = 5.6$eV.
The circles show our numerical data and the squares (connected by a full line) are the data points obtained from the expression \eqref{lqb}. Bottom panel: The same calculations as in the top panel, but $W$ is fixed to $9$nm and the Fermi energy is varied. To explore the small $\Delta$ limit reliably, we have to use the values of $\Delta$ which are too small to be realistic.} \label{lnmVSef}
\end{figure}

In the figure \ref{lnmVSef} the expression \eqref{lqb} is compared with $l_{qb}$ determined numerically by means of the approach discussed in figure \ref{Fig-55} (the right panel and inset to the right panel). The formula
\eqref{lqb} agrees with our numerical data if the roughness amplitude $\Delta$ is small. This is what one expects, because the perturbation expression \eqref{lqb} is exact in the limit $\Delta \rightarrow 0$ and our scattering matrix calculation is (in principle) exact for any $\Delta$.
As $\Delta$ increases, the result \eqref{lqb} fails to agree with our numerical data because the scattering is not weak \cite{comment4}.

Simple formulae can be derived for $l_{qb}$ and $l$ if the roughness is uncorrelated ($\Delta x / \lambda_F \ll 1$).
We start with $l$. If  $\Delta x / \lambda_F \ll 1$, the correlation function \eqref{Fq} is simply $\mathcal{F}(q) = \Delta x$. Consequently, the backscattering and forward-scattering terms in \eqref{Ker} become the same and the matrix \eqref{Ker} reduces to the diagonal form
\begin{eqnarray}
K_{n n'} = \delta_{n n'} \frac{2 \pi^2 \Delta^2 \Delta x}{3W^6}  \sum_{\mu=1}^{N_c} n^2 \mu^2 \frac{k_n}{k_{\mu}},
 \label{Kerdi}
\end{eqnarray}
where the factor of $2$ is just due the equal contribution of the backward and forward scattering.
We set \eqref{Kerdi} into the Boltzmann conductivity \eqref{sigmaB} and we extract from \eqref{sigmaB}
the diffusive mean free path. It reads
\begin{eqnarray}
l = \frac{3W^5}{\pi^4 \Delta^2 k_F \Delta x}  \sum_{n=1}^{N_c} \frac{k_n}{n^2} \left[ \sum_{\mu=1}^{N_c} \frac{\mu^2}{k_{\mu}} \right]^{-1}.
 \label{lunco}
\end{eqnarray}
For $N_c \gg 1$ the summations in \eqref{lunco} can be approximated as
\begin{eqnarray}
\sum_{n=1}^{N_c} \frac{k_n}{n^2} \simeq \frac{\pi^2}{6}k_F, \ \   \sum_{\mu=1}^{N_c} \frac{\mu^2}{k_{\mu}} \simeq \frac{W^3}{4 \pi^2} k^2_F,
 \label{sumaces}
\end{eqnarray}
and the semiclassical diffusive mean-free path becomes
\begin{eqnarray}
l = \frac{2 W^2}{\Delta^2 k^2_F \Delta x} .
 \label{luncofin}
\end{eqnarray}
Now we evaluate for $\Delta x / \lambda_F \ll 1$ the quasi-ballistic mean free path. We rewrite \eqref{lqb}
into the form
\begin{equation}
 l_{qb} = \frac{2}{\pi} N_c \left[ \sum_{m=1}^{N_c} \sum_{n=1}^{N_c} \frac{1}{l^R_{mn}} \right]^{-1}.
 \label{lqblmn}
\end{equation}
where
\begin{equation}
\frac{1}{l^R_{mn}} = \frac{2 \Delta^2 \kappa_m^2 \kappa_n^2}{3 W^2 k_m k_n} \mathcal{F}(|k_n + k_m|).
 \label{lqblmn help}
\end{equation}
Setting into \eqref{lqblmn help} the formula $\mathcal{F}(q) = \Delta x$ we obtain
\begin{equation}
l^R_{mn} = \frac{3W^2 k_m k_n}{2 \Delta^2 \kappa_m^2 \kappa_n^2 \Delta x} = \frac{3W^6 k_m k_n}{2 \pi^4 \Delta^2 m^2 n^2 \Delta x}.
\label{lRmn}
\end{equation}
Combining \eqref{lqblmn} with \eqref{lRmn} and using \eqref{sumaces} we obtain
\begin{equation}
  l_{qb} = \frac{2}{\pi} \frac{24 W}{\pi \Delta^2 k_F^3 \Delta x}.
 \label{lqbfin}
\end{equation}
We recall that $l_{qb}$ is limited exclusively by backscattering.

\begin{figure}[t!]
\begin{center}
\includegraphics[clip,width=0.9\columnwidth]{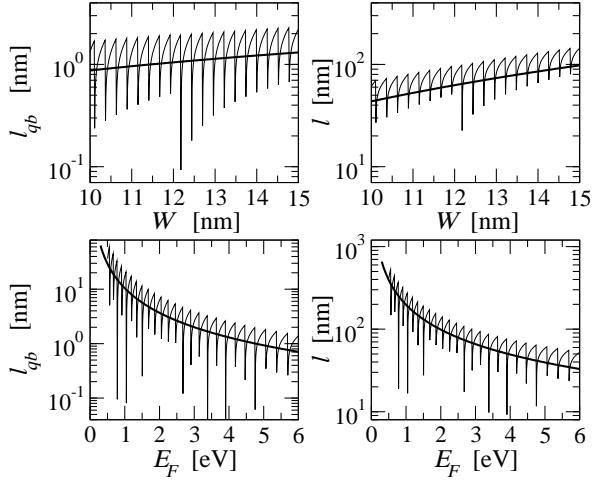}
\end{center}
\caption{Quasi-ballistic mean free path $l_{qb}$ and diffusive mean free path $l$ in the Au wire with rough edges, calculated as a function of the wire width $W$ and Fermi energy $E_F$. The thick lines show the formulae \eqref{lqbfin} and \eqref{luncofin}, derived for the uncorrelated roughness.
The thin lines show the formulae valid for an arbitrary correlation length $\Delta x$, namely the quasi-ballistic result
 \eqref{lqb} and the semiclassical $l$ extracted from the Boltzmann conductivity \eqref{sigmaB}.
 The roughness amplitude $\Delta$ and roughness correlation length $\Delta x$ are fixed to $\Delta = 0.5$nm and $\Delta x = 0.125$nm (the limit $\Delta x/\lambda_F \ll 1$ is fulfilled for all considered data).
 The Fermi energy used in the top panels is $E_F = 5.6$eV, the wire width in the bottom panels is $W = 9nm$. The oscillations with
 sharp minima appear whenever the Fermi energy approaches the bottom of the energy subband $n=N_c$.
} \label{lnmVSef1}
\end{figure}
The formulae \eqref{luncofin} and \eqref{lqbfin} hold for $\Delta x / \lambda_F \ll 1$. In figure \ref{lnmVSef1} we compare them with the original formulae valid for any $\Delta x$. We can see that the major difference is absence of the oscillating behavior in the formulae \eqref{luncofin} and \eqref{lqbfin}.

\begin{figure}[t]
\begin{center}
\includegraphics[clip,width=0.9\columnwidth]{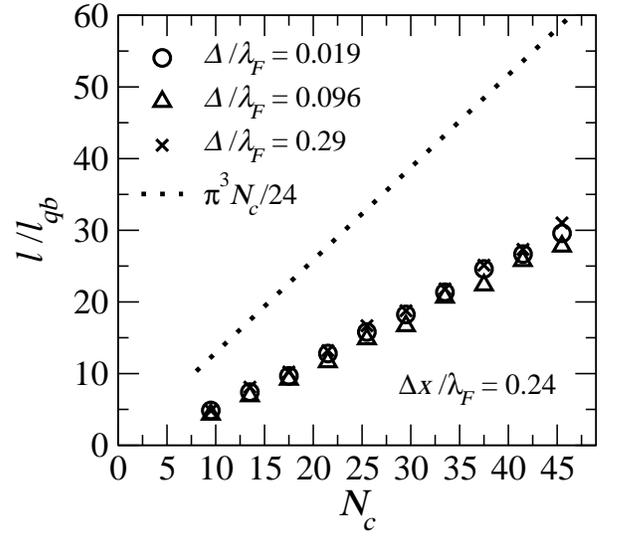}
\end{center}
\caption{Ratio $l/l_{qb}$, where $l$ is the diffusive mean free path and $l_{qb}$ is the quasi-ballistic mean free path, evaluated as a function of $N_c$ for various roughness amplitudes $\Delta$. The roughness correlation length $\Delta x$ is fixed to the value $\Delta x/\lambda_F = 0.24$ which reasonably emulates the uncorrelated roughness. The open symbols show the results of our quantum transport simulation, where
$l$ and $l_{qb}$ are calculated as in the figure \ref{Fig-55}.
} \label{l pomer lqb}
\end{figure}

Finally, the ratio $l/l_{qb}$ can be expressed as
\begin{equation}
\frac{l}{l_{qb}} = \frac{\pi^3}{24}N_c \sim N_c.
\label{ratio l lqb}
\end{equation}
 The result \eqref{ratio l lqb} is universal - independent on the wire parameters and parameters of disorder. The figure \ref{l pomer lqb} shows that the universal relation $l/\l_{qb} \propto N_c$ is confirmed by our exact quantum-transport calculation. Obviously, since the formula \eqref{luncofin} is the semiclassical Boltzmann-equation result and formula \eqref{lqbfin} is the weak-scattering limit, the formula \eqref{ratio l lqb} cannot reproduce the exact quantum results quantitatively.

The diffusive mean-free path \eqref{luncofin} contains both the backward and forward scattering. It is therefore interesting, that
the same result can be obtained when the quasi-ballistic (backward-scattering-limited)
resistance \eqref{Rqb} is extrapolated into the diffusive regime by means of the ansatz \eqref{Tnkadif}.
We start from
\begin{equation}
 \langle \rho \rangle \simeq 1/ \langle g \rangle =  \left[ \sum_{n=1}^{N_c}  \langle T_n \rangle \right]^{-1}
 \label{Rdifkot}
\end{equation}
and we use the ansatz \eqref{Tnkadif}. In the quasi-ballistic limit ($L \ll L_n$) we obtain from \eqref{Tnkadif} the formula $\langle T_n \rangle = 1 - L/L_n$. We set this formula into \eqref{Rdifkot} and we compare \eqref{Rdifkot} with the quasi-ballistic expressions \eqref{Rqb} and \eqref{lqblmn}. We find that
\begin{equation}
L_n = \left[ \sum_{m=1}^{N_c} \frac{1}{l^R_{mn}} \right]^{-1}.
\label{elenka}
\end{equation}
In the diffusive limit ($L_n \ll L$) we obtain from \eqref{Tnkadif} the formula $\langle T_n \rangle = L_n/L$ and from \eqref{Rdifkot}
the diffusive expression
\begin{equation}
 \langle \rho \rangle = \left[ \sum_{n=1}^{N_c}  L_n \right]^{-1}L.
 \label{Rdifkolim}
\end{equation}
We compare this expression with $\rho = \rho_{dif} \frac{L}{W}$,
where $\rho_{dif} = 2/(k_F l)$. We find the mean free path
\begin{equation}
l = \frac{2}{\pi N_c} \sum_{n=1}^{N_c}  L_n =  \frac{2}{\pi N_c} \sum_{n=1}^{N_c}  \left[ \sum_{m=1}^{N_c} \frac{1}{l^R_{mn}} \right]^{-1}.
\label{ldiflmn}
\end{equation}
If we set into \eqref{ldiflmn} the uncorrelated limit \eqref{lRmn}, we obtain again the Boltzmann mean-free path \eqref{luncofin}.
This is the proof that the ansatz \eqref{Tnkadif} works correctly for the uncorrelated roughness.
Now it is useful to make two remarks.

First, our characteristic length $L_n$ should not be confused with the often used \cite{Izrailev,Izrailev2} attenuation length $L_n$.
Our $L_n$ is defined by the ansatz \eqref{Tnkadif} and we have just seen, that the expression \eqref{ldiflmn} gives for such $L_n$
the mean free path coinciding with the mean free path obtained from the Boltzmann equation. This is the momentum-relaxation-time-limited mean-free path. However, if one sets into \eqref{ldiflmn} the attenuation length (equation (5.2) in \cite{Izrailev2}), one obtains from \eqref{ldiflmn} the scattering-time-limited mean-free path, i.e., the mean distance between two subsequent collisions. For the uncorrelated roughness the latter is exactly twice shorter than the former one.

Second, we note that the ansatz \eqref{Tnkadif} does not work for arbitrary roughness. Indeed, we can set into the formula \eqref{ldiflmn} the more general expression \eqref{lqblmn help} and we can compare \eqref{ldiflmn} with the Boltzmann mean free path
valid for an arbitrary correlation length $\Delta x$. In such case the formula \eqref{ldiflmn} fails to reproduce the Boltzmann-equation result.

Assuming the uncorrelated roughness, we are ready to derive analytically the effective number of the open channels, $N^{eff}_c$. The expression \eqref{Rdifkot} can be formally written as
\begin{equation}
 \langle \rho \rangle = \frac{1}{N^{eff}_c}  + \frac{2}{k_F l} \frac{L}{W},
 \label{Rdifkolimeff}
\end{equation}
where the symbol $N^{eff}_c$ is defined as
\begin{equation}
  N^{eff}_c = \left[ \left( \sum_{n=1}^{N_c}  \langle T_n \rangle \right)^{-1} - \frac{2}{\pi N_c l} L \right]^{-1}
 \label{Neff1}
\end{equation}
in order to obtain \eqref{Rdifkot} again.
We now express the transmission $ \langle T_n \rangle$ by means of the formula \eqref{Tnkadif} and the mean free path $l$ by means
of the formula \eqref{ldiflmn}. We obtain
\begin{equation}
 N^{eff}_c = \left[ \left( \sum_{n=1}^{N_c}  \frac{L_n}{L_n + L} \right)^{-1} - \left( \sum_{n=1}^{N_c}  L_n \right)^{-1} L \right]^{-1}.
 \label{Neff2}
\end{equation}
In the diffusive limit ($L \gg L_n$) we obtain after some algebraic manipulations the equation
\begin{equation}
 N^{eff}_c = \frac{\left( \sum_{n=1}^{N_c}  L_n \right)^{2}}{\sum_{n=1}^{N_c}  L^2_n}.
 \label{Neffin}
\end{equation}
The expression \eqref{Neffin} no longer depends on $L$ and evidently represents the effective number of the open channels, if the resistance \eqref{Rdifkolimeff} is considered
in the diffusive limit.
We set into  \eqref{Neffin} the formulae \eqref{elenka} and \eqref{lRmn}. We obtain
\begin{equation}
  N^{eff}_c= \frac{\left[ \sum_{n=1}^{N_c} \frac{k_n}{n^2} \right]^2 \left[ \sum_{m=1}^{N_c} \frac{m^2}{k_m} \right]^{-2}}{\left[ \sum_{n=1}^{N_c} \frac{k^2_n}{n^4} \right] \left[ \sum_{m=1}^{N_c} \frac{m^2}{k_m} \right]^{-2}}.
 \label{Neffin1}
\end{equation}
For $N_c \gg 1$ the first sum in the denominator of \eqref{Neffin1} becomes
\begin{equation}
 \sum_{n=1}^{N_c} \frac{k^2_n}{n^4} \simeq \frac{\pi^4}{90}k^2_F
 \label{sumneff}
\end{equation}
and other sums in \eqref{Neffin1} are already known [see equations \eqref{sumaces}].
We arrive to the result
\begin{equation}
 N^{eff}_c \approx \frac{5}{2},
 \label{Neffuni}
\end{equation}
which implies that $N^{eff}_c$ is a universal number depending neither on the roughness amplitude $\Delta$ nor on the number of the conducting channels, $N_c$. All this agrees with our quantum-transport calculation in the figure \ref{Fig-44}, except that the result \eqref{Neffuni} underestimates the numerical
value $N^{eff}_c \simeq 6$ about twice. We however recall that the result \eqref{Neffuni} relies on the formulae \eqref{elenka} and \eqref{lRmn} which are not exact.

\subsection{D. Diffusive mean free path: quantum-transport results versus the semiclassical Boltzmann results}

In this subsection, the mean free path
 obtained from the semiclassical Boltzmann equation is compared with the mean free path obtained from the quantum resistivity $\rho_{dif}$. For the impurity disorder we see a standard result \cite{cahay,tamura}: the semiclassical and quantum mean-free paths coincide for weak impurities. On the contrary, for the edge roughness we find, that the quantum mean-free path differs from the semiclassical one
even if the roughness amplitude $\Delta$ is vanishingly small.

\begin{figure}[t]
\begin{center}
\includegraphics[clip,width=0.9\columnwidth]{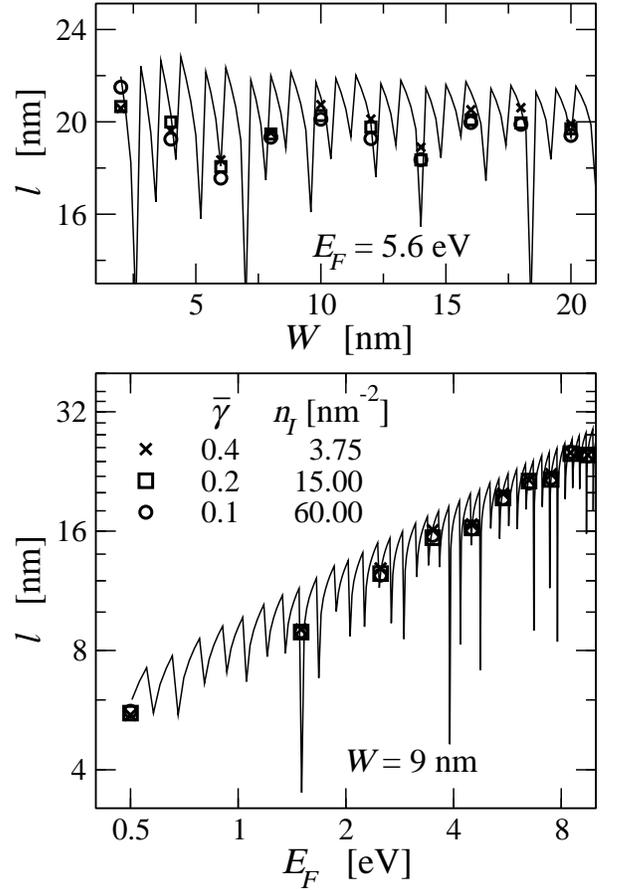}
\end{center}
\caption{Top panel: Diffusive mean free path $l$ in the Au wire with impurity disorder as a function of the wire width $W$, calculated for three different sets of the impurity parameters (listed in the bottom panel). The semiclassical mean free path obtained from the Boltzmann Q1D conductivity \eqref{sigmaIMPs} is shown in a full line: all three sets of parameters are intentionally chosen to give the same semiclassical result.
The data shown by symbols are the quantum-transport results obtained from the quantum resistivity $\rho_{dif}$ (c.f. figure \ref{Fig-55}). Bottom panel: The same calculations as in the top panel, but $W$ is fixed to $9$nm and the Fermi energy is varied. 
} \label{Fig-9}
\end{figure}

In the figure \ref{Fig-9} the quantum and semiclassical mean free paths are compared for the impurity disorder.
The semiclassical data are shown in a full line: three different sets of parameters are intentionally chosen to provide the same semiclassical result.
The full lines exhibit oscillations with sharp minima, appearing whenever the Fermi energy approaches the bottom of the energy subband $n=N_c$. Evidently,
the oscillating curve intersects the quantum results (the data shown by symbols), albeit they are slightly affected by statistical noise. Both the semiclassical and quantum results follow the trend predicted by the semiclassical 2D limit (equation  \ref{sigmaIMP}), namely that $l$ is proportional to $E^{1/2}_F$ and independent on $W$. In summary, in the Q1D wire with impurity disorder the semiclassical and quantum mean-free paths coincide. This is a standard result \cite{cahay,tamura}, known from the theory based on the white-noise disorder \cite{Mello-book} (see however \cite{Mosko}).

\begin{figure}[t]
\begin{center}
\includegraphics[clip,width=0.9\columnwidth]{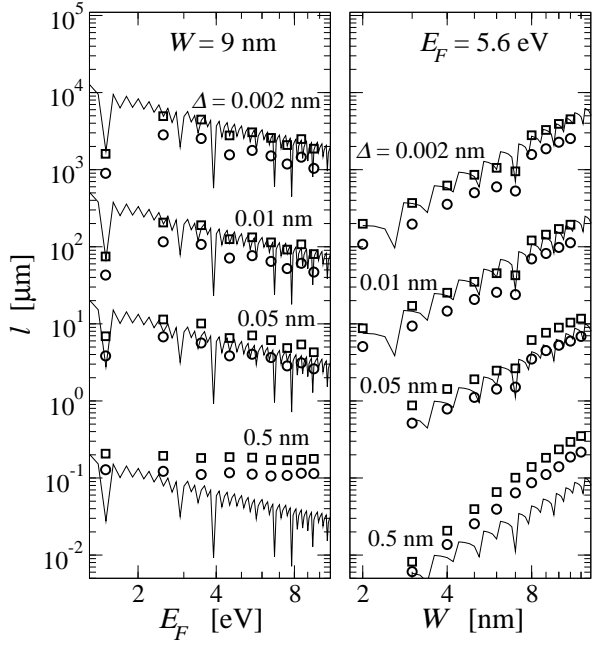}
\end{center}
\caption{The right panel shows the diffusive mean free path $l$ in the Au wire with rough edges as a function of the wire width $W$, calculated for various roughness amplitudes $\Delta$. The left panel shows the same calculation, but $W$ is fixed to $9$nm and the Fermi energy is varied. The roughness correlation length is kept at the value $\Delta x = 0.125$nm, which means that $\Delta x/\lambda_F \lesssim 1$ for most of the presented data. The full lines show the semiclassical mean free path obtained from the Boltzmann Q1D conductivity (equations \ref{sigmaB}
and \ref{Ker}). The open circles show the quantum mean free path extracted from the quantum resistivity $\rho_{dif}$ (c.f. figure \ref{Fig-55}). The open squares show the mean free path obtained by means of the classical scattering-matrix calculation (see the text), in which the localization is absent. To explore the small $\Delta$ limit reliably, we have to use the values of $\Delta$ which are too small to be realistic.
} \label{Fig-934}
\end{figure}

As shown in the figure \ref{Fig-934}, in the wire with rough edges the situation is different.
We again compare  the semiclassical mean free path (full lines) with the quantum mean free path (open circles).
As before, the full lines exhibit oscillations with
sharp minima, appearing whenever the Fermi energy approaches the bottom of the energy subband $n=N_c$.
   Evidently, the full line and open circles show a quite different behavior if the roughness amplitude $\Delta$ is large (notice the results for $\Delta = 0.5$nm). It is tempting to ascribe this difference to the weak-perturbation approximation involved in the Boltzmann equation, and similarly, it is tempting to expect that the full line will intersect the open circles for sufficiently small $\Delta$. However, this is not the case: even for the smallest considered $\Delta$ the open circles are systematically a factor of $\sim 2$ below the full line. In summary, in the wire with rough edges the quantum mean-free path differs from the semiclassical one (by the factor of $\sim 2$) even if $\Delta$ is vanishingly small.

   To understand the origin of this difference, we now exclude from our quantum-transport calculation the wave interference.
We recall (see subsection II.C), that in the quantum-transport calculation the total $S$-matrix of the disordered wire is obtained by combining at random the scattering matrices ($b_i$) of the building blocks, where the $2N \times 2N$ matrix $b_i$ is composed of the complex amplitudes $t_{mn}, t'_{mn}, r_{mn}$, and $r'_{mn}$. To exclude the wave interference, we proceed as follows \cite{cahay}. First, we consider only the conducting channels and we completely neglect the evanescent ones: this reduces the size of the matrix $b_i$ to $2N_c \times 2N_c$. Second, instead of the complex matrix $b_i$ we use the real one, in which
the complex amplitudes  $t_{mn},t'_{mn},r_{mn}$, and $r'_{mn}$ are replaced by the real probabilities $T_{mn} = |t_{mn}|^2 ,T'_{mn} = |t'_{mn}|^2,R_{mn} = |r_{mn}|^2$, and $R'_{mn} = |r'_{mn}|^2$, respectively. Of course, the wave interference is excluded completely, if the length of the  building block, $L_b$, coincides with the length of the edge step, $\Delta x$. Fortunately, in practical calculations the wave interference is negligible
already for $L_b \sim l$. If we combine the resulting real matrices $b_i$ by means of the same combination law as before (equation \ref{skladka}), we obtain the classical transmission probability $\mathcal{T}_{mn}$ and eventually the classical Landauer conductance
\begin{equation}
g_{clas} = \sum_{m=1}^{N_c} \sum_{n=1}^{N_c} \mathcal{T}_{mn} \frac{k_m}{k_n}.
\end{equation}
Finally, we perform ensemble averaging over many samples.

\begin{figure}[t]
\centerline{\includegraphics[clip,width=\columnwidth]{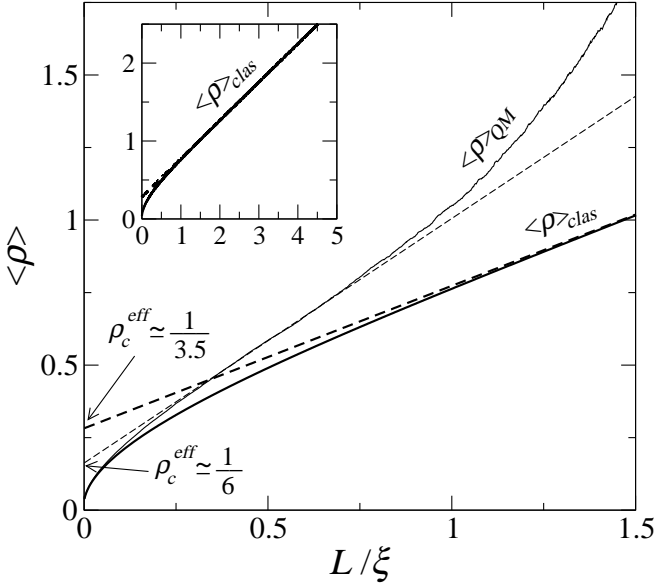}}
\vspace{-0.15cm} \caption{Mean resistance $\langle \rho \rangle$ of the wire with rough edges in dependence on the ratio $L/\xi$. The thin full line  is the result of the quantum scattering-matrix calculation (the same data as in the top right panel of figure \ref{Fig-3}), now labeled as ${\langle \rho \rangle}_{QM}$. The thick full line is the result of the classical scattering-matrix calculation, labeled as ${\langle \rho \rangle}_{clas}$.
The dashed lines show the linear fits $\langle \rho \rangle = \rho^{eff}_c + \rho_{dif}L/W$, from which we determine the effective contact resistance $\rho^{eff}_c$ and the diffusive mean free path $l = 2/(k_F \rho_{dif})$. The resulting $\rho^{eff}_c$ are shown in the figure. Inset shows the classical result again, but for larger $L/\xi$ in order to stress the linear raise and absence of the localization. } \label{Fig-8}
\end{figure}
In the figure \ref{Fig-8} the classical scattering-matrix calculation is compared with the quantum one. The mean resistance due to the quantum calculation is labeled as ${\langle \rho \rangle}_{QM}$ to distinguish from the classical ${\langle \rho \rangle}_{clas}$.
It can be seen that both ${\langle \rho \rangle}_{QM}$ and ${\langle \rho \rangle}_{clas}$ exhibit the crossover to the linear diffusive dependence
$\langle \rho \rangle = \rho^{eff}_c + \rho_{dif}L/W$, but the classical result shows a smaller value of $\rho_{dif}$ and a larger value of $\rho^{eff}_c$. The value $\rho^{eff}_c = 1/N^{eff}_c \simeq 1/3.5$
is already close to our theoretical result $N^{eff}_c \simeq 2.5$ (equation \ref{Neffuni}), where the wave interference is excluded as well.
The smaller value of $\rho_{dif}$ is due to the absence of the localization.

 Let us return to the figure \ref{Fig-934}. The squares show the mean free path extracted from ${\langle \rho \rangle}_{clas}$. These data overestimate the quantum results (circles) systematically by a factor of $\sim 2$. Compare now the classical scattering-matrix results (squares) with the semiclassical Boltzmann results (full lines).
 The full lines intersect the squares only if $\Delta$ is vanishingly small.
  This is due to the fact that the Boltzmann results rely on the weak-perturbation theory while the classical scattering-matrix calculation still involves the exact quantum scattering by a single edge step.

The roughness-limited conductivity derived from the Boltzmann equation (usually further simplified by using the Fuchs-Sondheimer model \cite{Fuchs2,Sondheimer}) is often compared with the transport measurements of the metallic nanowires \cite{Durkan,Steinhogel,Graham}. However, such comparison
 cannot verify the Boltzmann result unambiguously due to the presence of other processes (like the scattering by impurities and grain boundaries) and also due to the fact, that the roughness parameters $\Delta$ and $\Delta x$ are not known \emph{apriori}. Our quantum calculation shows unambiguously, that the semiclassical Boltzmann approach describes the wires with rough edges reasonably, only if $\Delta$ \emph{is too small to be realistic (in the Au nanowires)}.
 In practice, the metallic nano-wires are usually fabricated by advanced lift-off techniques \cite{suklee,voigt,bryan}, which hardly allow to suppress the edge roughness to the value $\Delta \sim 1$nm.

 In the table \ref{tabula}, the quantum and semiclassical mean-free paths are compared for a more realistic $\Delta$ and $W$ like in the figure \ref{Fig-934}. We see that the quantum result can exceed the semiclassical Boltzmann result one order of magnitude. One might argue that the quantum result holds only in the coherent regime while decoherence is present at any nonzero temperature. We expect that decoherence will tend to destroy the wave interference and to restore the resistance ${\langle \rho \rangle}_{clas}$ resulting from the classical scattering-matrix approach. The resulting classical mean-free path would be twice larger than the quantum one [see figure \ref{Fig-934}], i.e., it would exceeds the Boltzmann mean-free path in the table \ref{tabula} by another factor of two.

\begin{table}[t!]
\begin{tabular}{|c|c|c|c|c|}
	\hline
$W$[nm] &  $\Delta x$ [nm]  &   $\Delta$ [nm]  & $l$ [nm]   & $l_{Boltz}$ [nm]  \\
	\hline
9   & 0.5 & 1.5 & 21.4  & 5.17 \\
20   & 10 & 3.33 & 70.7  & 21.3\\
20   & 5 & 3 & 65.6 & 17.2 \\
30   & 2.5 & 4 & 124  & 14.8 \\
30   & 5 & 4.5 & 105  & 16.6 \\
70   & 5 & 10.5 & 315  & 20.5 \\
90   & 5 & 13.5 & 409  & 30.0 \\
90   & 40 & 13.5 & 411  & 115 \\
	\hline
\end{tabular}
\caption{The mean-free path $l$ in the Au wire obtained from the quantum-transport simulation is compared with the semiclassical Boltzmann mean-free path $l_{Boltz}$ (equations \ref{sigmaB} and \ref{Ker}) for various values of the wire width $W$, roughness amplitude $\Delta$ and roughness-correlation length $\Delta x$. Each set of the parameters considered in the table represents strong disorder, for which $l_{Boltz}$ strongly differs from $l$.}
\label{tabula}
\end{table}

 \subsection{E. Wires with strongly correlated edge roughness: $\Delta x/\lambda_F \ggg 1$}

 The wires with rough edges have sofar been discussed for the roughness-correlation lengths $\Delta x \lesssim \lambda_F$.
 The only exception was the figure \ref{Fig-44} where the effective number of the open channels, $N^{eff}_c$, was calculated from $\Delta x \ll \lambda_F$ up to $\Delta x \ggg \lambda_F$. We have seen, that the value of $N^{eff}_c$ approaches for $\Delta x \ggg \lambda_F$  the value of $N_c$, which suggests that the diffusive dependence $\langle \rho \rangle = 1/N^{eff}_{c} + \rho_{dif} L/W$ approaches the standard
 form $\langle \rho \rangle = 1/N_{c} + \rho_{dif} L/W$, seen in the wire with impurities. In this subsection we focus on the limit $\Delta x \ggg \lambda_F$.

 \begin{figure}[t!]
\centerline{\includegraphics[clip,width=\columnwidth]{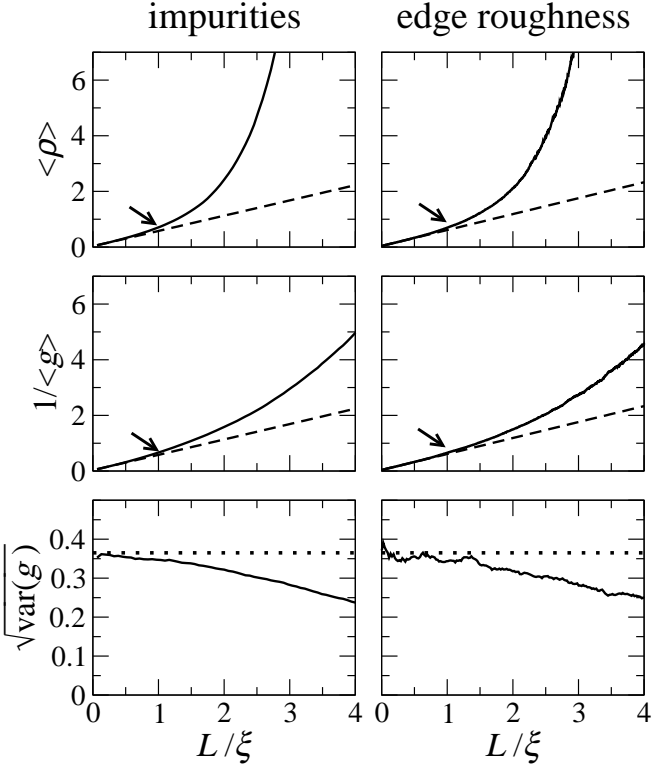}}
\vspace{-0.15cm} \caption{The same calculation and the same symbols as in the figure \ref{Fig-3}, except that the edge roughness is considered in the limit $\Delta x/\lambda_F \ggg 1$. Specifically, the data obtained for the wire with impurities (left column) are the same as those in the left column of figure \ref{Fig-3}, while for the edge roughness (right column) we show the data obtained for $\Delta x/\lambda_F = 19300$ and $\Delta/\lambda_F=0.096$ (the data for $\Delta/\lambda_F=0.019$ and $\Delta/\lambda_F=0.96$, not shown, look similarly except for statistical noise).
We recall that the localization length $\xi$ is determined from the dependence $\langle \ln{g} \rangle = - L/\xi$. For the edge roughness
we now obtain $\xi/l \simeq 0.92 N_c$, which is close to the impurity case: $\xi/l \simeq 0.88 N_c$
} \label{Fig-323}
\end{figure}
 The figure \ref{Fig-323} shows the same calculations as the figure \ref{Fig-3}, except that the wire with rough edges is now simulated in the limit $\Delta x \ggg \lambda_F$. Unlike the figure \ref{Fig-3},
 the results for the wire with rough edges are now close to the results for the wire with impurity disorder.

 First, the localization length in the wire with rough edges fulfills the relation $\xi/l \simeq 0.92 N_c$, which is close to the relation $\xi/l \simeq 0.88 N_c$ found for the impurity disorder. This is why the resistance $\langle \rho \rangle$ and inverse conductance $1/\langle g \rangle$ in the figure \ref{Fig-323} show for both types of disorder the same linear slope with $L/\xi$. Moreover, the same linear slope means a single linear regime for the quasi-ballistic as well as diffusive transport, i.e., the crossover between the quasi-ballistic and diffusive regimes, observed for $\Delta x \lesssim \lambda_F$, tends to disappear. (In the figure \ref{Fig-3} a remarkably larger slope is seen for the edge roughness because $\xi/l \simeq 1.4 N_c$.)

 Second, the inverse conductance in the figure \ref{Fig-323} shows for both types of disorder the linear diffusive regime for $L/\xi < 1$ and onset of localization for $L/\xi \simeq 1$. (In the figure \ref{Fig-3} the inverse conductance of the wire with rough edges shows onset of localization for $L/\xi \simeq 2$.)

 Third, the wire with rough edges now exhibits essentially the same universal conductance fluctuations as the wire with impurities. The size of these fluctuations is in accord with the value $\sqrt{\mbox{var}(g)} = 0.365$, predicted for the white-noise-like disorder.
 To see such conductance fluctuations for the edge roughness is surprising: sofar only the value $\sqrt{\mbox{var}(g)} \simeq 0.3$ has been reported \cite{MartinSaenz2,Nikolic,SanchezGil,AndoTamura}.

\begin{figure}[t]
\centerline{\includegraphics[clip,width=\columnwidth]{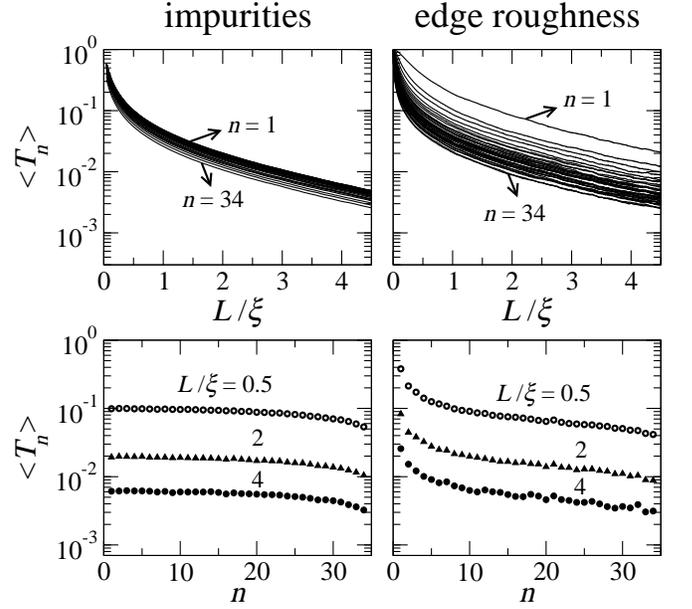}}
\vspace{-0.15cm} \caption{The same calculation and the same symbols as in the figure \ref{Fig-4}, except that the edge roughness is considered in the limit $\Delta x/\lambda_F \ggg 1$ [we show the data for $\Delta x/\lambda_F = 19300$].
} \label{Fig-4434}
\end{figure}
 To provide insight,
 it is useful to look at the transmission probabilities  $\langle T_n \rangle$.
 The figure \ref{Fig-4434} shows the same calculation as the figure \ref{Fig-4}, but with the edge roughness considered in the limit $\Delta x/\lambda_F \ggg 1$. It can be seen that the figure \ref{Fig-4434} differs from the figure \ref{Fig-4} in the following respect: except for a few channels with lowest values of $n$, the wire with rough edges and wire with impurities show
 for the rest of the channels a very similar $\langle T_n \rangle$, which is clearly not the case in figure \ref{Fig-4}. This similarity
 is responsible for similarity of the transport results in the figure \ref{Fig-323}. The observed similarity could be further improved by simulating the wires with $N_c$ larger than $34$, but this is beyond the scope of this work.

\begin{figure}[t]
\begin{center}
\includegraphics[clip,width=\columnwidth]{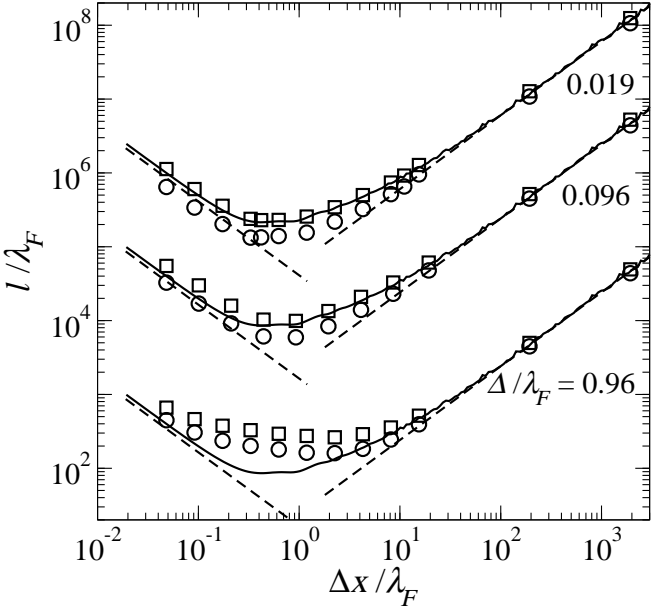}
\end{center}
\caption{
Diffusive mean-free path $l$ in the wire with rough edges as a function of the roughness-correlation length $\Delta x$
for various roughness amplitudes $\Delta$, with $l$, $\Delta x$ and $\Delta$ scaled by $\lambda_F$. The wire width $W$ is fixed
to the value $W/\lambda_F = 17.4$, which means that $N_c = 34$). The full lines show the semiclassical mean-free path obtained from the Boltzmann conductivity (equations \ref{sigmaB}
and \ref{Ker}). The dashed lines for $\Delta x/\lambda_F \lesssim 1$ show the uncorrelated limit \eqref{luncofin}, the
dashed lines for $\Delta x/\lambda_F \gtrsim 1$ show the correlated limit \eqref{sigmaBsl}. The circles show the quantum mean-free path extracted from the quantum resistivity $\rho_{dif}$ (c.f. figure \ref{Fig-55}).
The squares show the mean-free path obtained by means of the classical scattering-matrix calculation in which the localization effect is absent (see subsection IV.D). Note that for the Au wire ($E_F = 5.6eV$ and $m=9.109\times10^{-31}$kg) we have $\lambda_F = 0.52$nm, i.e., the wire width is $W = 17.4 \lambda_F = 9$nm and the smallest $\Delta$ is only $0.01$nm. Such small $\Delta$ is obviously not realistic, which demonstrates an important finding: The semiclassical Boltzmann result (full line) reproduces the classical scattering-matrix calculation (squares) for any $\Delta x$ only if $\Delta$ is too small to be realistic.
} \label{Fig-13}
\end{figure}

 Finally, the figure \ref{Fig-13} shows for $\Delta x \ggg \lambda_F$, what happens in the wire with rough edges with the diffusive mean free path.
 Specifically, the semiclassical Boltzmann result (full lines) is compared with the quantum (open circles) and classical (open squares) scattering-matrix calculations. In the limit $\Delta x \ggg \lambda_F$ all three calculations tend to give the same mean free path. This means that
 the interference effects observed for $\Delta x \lesssim \lambda_F$ are no longer important, or in other words, the diffusive transport in the wire with rough edges is semiclassical. This is again similar to the wire with impurities.

For $\Delta x \ggg \lambda_F$ it is easy to show analytically, that the semiclassical Boltzmann mean free path
 rises with $\Delta x$ linearly. The correlation function $\mathcal{F}(q)$ contains the function $\cos(\Delta x q)$. This function oscillates
 with $q$ at random and the oscillations become very fast for $\Delta x \ggg \lambda_F$. Consequently,
 the correlation function $\mathcal{F}(q)$ oscillates fast around the value
$ \mathcal{F}(q) = \frac{2}{q^2\Delta x }$.
If we ignore these fast oscillations, the diffusive mean free path can be written as
\begin{equation}
 l = \frac{3 \Delta x W^5}{k_F\pi^4 \Delta^2} \sum^{N_c}_{n} \sum^{N_c}_{n'} k_n k_{n'} (\mathcal{K})^{-1}_{n n'}
 \label{sigmaBsl}
\end{equation}
where
\begin{eqnarray}
\mathcal{K}_{n n'} = \delta_{n n'} \left[ \sum_{\mu \neq n}^{N_c} n^2 \mu^2
\frac{k_n}{k_{\mu}} \left[
\frac{1}{(k_{\mu}-k_n)^2} + \frac{1}{(k_{\mu}+k_n)^2} \right] \right.
\nonumber \\
 + \frac{n^4}{2k_n^2} \Bigg] -  (1 - \delta_{n n'}) n^2 {n'}^2 \left[
\frac{1}{(k_{n'}-k_n)^2} - \frac{1}{(k_{n'}+k_n)^2} \right]. \nonumber \\
 \label{Kersl}
\end{eqnarray}
It can be seen that $l \propto \Delta x$. The $l \propto \Delta x$ dependence resembles the wire with impurities, where $l$ rises linearly with the average distance between the impurities. The formula \eqref{sigmaBsl} is plotted in the figure \ref{Fig-13} in a dashed line. It indeed agrees with the numerical data for $\Delta x \ggg \lambda_F$. On the other hand, the formula \eqref{luncofin}, derived in the opposite limit, reasonably agrees with the numerical data for $\Delta x \ll \lambda_F$, but the dependence $l \propto 1/\Delta x$ has no similarity with the impurity case.

Finally, as also pointed out in the figure caption, the figure \ref{Fig-13} shows that the semiclassical Boltzmann result (full line) reproduces the classical scattering-matrix calculation (squares) for any $\Delta x$ only if $\Delta$ is too small to be realistic. This is another indication that the Boltzmann-equation theory of the edge-roughness-limited transport should be used with caution at least in the Au nanowires.

\section{V. Summary and concluding remarks}

We have studied quantum transport in quasi-one-dimensional wires
made of a two-dimensional conductor of width $W$ and length $L \gg W$. The main purpose of our work was to compare an impurity-free wire with rough edges with a smooth wire with impurity disorder. We have calculated the electron transmission through the wires by
the scattering-matrix method, and we have obtained the Landauer conductance/resistance for a large ensemble of macroscopically identical disordered wires.

We have first studied the impurity-free wire whose edges have a roughness correlation length comparable with the Fermi wave length.
The mean resistance $\langle \rho \rangle$ and inverse mean conductance $1/ \langle g \rangle$ have been evaluated in dependence on $L$.
For $L \rightarrow 0$
we have found the quasi-ballistic dependence $1/ \langle g \rangle = \langle \rho \rangle = 1/N_c + \rho_{qb} L/W$,
where $1/N_c$ is the fundamental
contact resistance and $\rho_{qb}$ is the quasi-ballistic resistivity. For larger $L$ we have found the crossover to the diffusive dependence $1/ \langle g \rangle \simeq \langle \rho \rangle = 1/N^{eff}_c + \rho_{dif} L/W$, where
$\rho_{dif}$ is the resistivity and
$1/N^{eff}_c$ is the effective contact resistance corresponding to the $N^{eff}_c$ open channels. We have found the universal results $\rho_{qb}/\rho_{dif} \simeq 0.6N_c$ and $N^{eff}_c \simeq 6$ for $N_c \gg 1$.

Approaching the localization regime, we have demonstrated the following numerical finding: as $L$ exceeds the localization length $\xi$,
the resistance shows onset of localization while the conductance shows the diffusive dependence $1/ \langle g \rangle \simeq 1/N^{eff}_c + \rho_{dif} L/W$ up to $L \simeq 2 \xi$ and the localization for $L > 2 \xi$ only. On the contrary, for the impurity disorder we have found a standard diffusive behavior, namely $1/ \langle g \rangle \simeq \langle \rho \rangle \simeq 1/N_c + \rho_{dif} L/W$ for $L < \xi$.
We have also seen that the impurity disorder and edge roughness show the universal conductance fluctuations of different size, as already reported
in previous works \cite{MartinSaenz2,Nikolic,SanchezGil,AndoTamura}.

We have then attempted to interpret our quantum-transport results in terms of an approximate but microscopic analytical theory.
In particular, the crossover from $1/ \langle g \rangle = \langle \rho \rangle = 1/N_c + \rho_{qb} L/W$ to $1/ \langle g \rangle \simeq \langle \rho \rangle = 1/N^{eff}_c + \rho_{dif} L/W$ in the wire with rough edges has been derived analytically assuming the uncorrelated edge roughness and neglecting the localization effects. In spite of this approximation, our analytical results capture the main features of our numerical results. Specifically, we have derived the universal results $\rho_{qb}/\rho_{dif} = \frac{\pi^3}{24}N_c$  and $N^{eff}_c \simeq 2.5$.

We have also derived the wire conductivity from the semiclassical Boltzmann equation, and we have compare the semiclassical
mean-free path
with the mean-free path obtained from the quantum resistivity $\rho_{dif}$. For the impurity disorder we have found, that the semiclassical and quantum mean-free paths coincide, which is a standard result. However, for the edge roughness the semiclassical mean-free path strongly differs from the quantum one, showing that the diffusive transport in the wire with rough edges is not semiclassical. We have found that it becomes semiclassical only if the roughness-correlation length is much larger than the Fermi wave length. In such case the resistance and conductance tend to scale with $L/\xi$ like in the wire with impurity disorder, also showing the universal conductance fluctuations of similar size.

We end by a remark about our edge-roughness model. It is the same model as in the previous simulations \cite{garcia,MartinSaenz2,feilhauer,MartinSaenz,MartinSaenz3,MartinSaenz4}. We believe that most of the results obtained within the model are model-independent. For example, we have tested the correlation function $\mathcal{F}(q)$
of Gaussian shape and we have seen that the resulting transmissions remain quite similar to those in the figure \eqref{Fig-4}
The only exception is the correlated limit \eqref{sigmaBsl}, which predicts the dependence $l \propto \Delta x$.
This dependence is the artefact of our $\mathcal{F}(q)$ choice. For example, for the Gaussian correlation function,
the dependence $l \propto \Delta x$ is replaced by a much faster increase with $\Delta x$.
On the other hand, the uncorrelated limit \eqref{luncofin} can be equally well derived for the Gaussian correlation function or for the exponential one, i.e., the result \eqref{luncofin} is very general.

\section{Acknowledgement}
We thank for the grant VVCE-0058-07, VEGA grant 2/0633/09, and grant APVV-51-003505.
We thank Peter Marko\v s for useful discussions in the initial stage of this work and for helpful comments to the manuscript.

\section{Appendix A: Derivation of the relaxation time  }

At zero temperature the distribution \eqref{froz 1} reads
\begin{equation}
 f_n(k) = f(E_{nk}) - \frac{eF \hbar}{m}k \tau_n(E_{nk})\delta(E_{nk} - E_F)
 \label{froz0 1}
\end{equation}
and the Boltzmann equation \eqref{Boltz lin} can be written as
\begin{eqnarray}
\frac{eF \hbar}{m}k \delta(E_{nk} - E_F) &=& \sum_{n'} \sum_{k'} W_{n,n'}(k,k') \nonumber \\
&\times& \left[ f_{n'}(k') - f_n(k) \right].
 \label{Boltz 1}
\end{eqnarray}
We set \eqref{froz0 1} and \eqref{Wborn 1} into \eqref{Boltz 1}. We obtain
\begin{eqnarray}
&&  k \delta(E_{nk} - E_F) = \sum_{n'} \sum_{k'} \frac{2 \pi}{\hbar} |\langle n' ,k' | U | n, k\rangle|^2_{\mbox{\fontsize{8}{8}\selectfont av}} \nonumber \\
&\times& \delta(E_{nk} - E_{n'k'}) \Big[ k'\tau_{n'}(E_{n'k'}) \delta(E_{n'k'} - E_F)  \nonumber \\
&-& k \tau_n(E_{nk})\delta(E_{nk} - E_F)\Big].
\label{Boltze dosadene}
\end{eqnarray}
We proceed similarly to the Q2D theory \cite{Fishman1}. First, we multiply both sides of \eqref{Boltze dosadene} by $k$ and sum over $k$. Second, on the left hand side we replace $\sum_k$ by $(L/2\pi) \int dk$ and integrate. Third, on the right hand side we
replace $\tau_n(E_{nk})\delta(E_{nk} - E_F)$ by $\tau_n(E_F)\delta(E_{nk} - E_F)$ and $\delta(E_{nk}-E_{n'k'})\delta(E_{nk} - E_F)$
by $\delta(E_{n'k'} - E_F)\delta(E_{nk} - E_F)$, which is justified due to the presence of $\delta(E_{nk} - E_F)$.
We obtain the equation
\begin{eqnarray}
\frac{mL}{\pi \hbar^2}k_n &=& \sum_{n'} \sum_{k} \sum_{k'} \frac{2 \pi}{\hbar} |\langle n ,k | U | n', k'\rangle|^2_{\mbox{\fontsize{8}{8}\selectfont av}} \delta(E_{nk}-E_F) \nonumber \\
&\times&  \delta(E_{n'k'} - E_F) \Big[ k^2\tau_{n}(E_F) - kk' \tau_{n'}(E_F)\Big]
\label{Boltzes 1}
\end{eqnarray}
and from $\eqref{Boltzes 1}$ eventually
\begin{equation}
\frac{m}{\pi \hbar^2}k_n = \frac{2 \pi}{\hbar} \sum_{n'} K_{nn'} \tau_{n'}(E_F),
\label{Boltzes 2}
\end{equation}
where $K_{nn'}$ is defined by equation \eqref{KmatkaA 1}. From \eqref{Boltzes 2} we obtain the relaxation time
\eqref{trelax}.

\section{Appendix B: Golden rule for scattering by impurity disorder}

If the scattering potential $U$ is given
by the impurity potential \eqref{delta}, then
\begin{equation}
|\langle n' ,k' | U | n, k\rangle|^2_{\mbox{\fontsize{8}{8}\selectfont av}} =  |\langle n' ,k' | \sum_{i=1}^{N_I} \gamma \delta(x-x_i) \delta(y-y_i) | n, k\rangle|^2_{\mbox{\fontsize{8}{8}\selectfont av}}.
 \label{rozptIMP}
\end{equation}
Since the impurities are positioned at random, equation  \eqref{rozptIMP} can be readily simplified as
\begin{equation}
|\langle n' ,k' | U | n, k\rangle|^2_{\mbox{\fontsize{8}{8}\selectfont av}} =  \sum_{i=1}^{N_I} |\langle n' ,k' |  \gamma \delta(x-x_i) \delta(y-y_i) | n, k\rangle|^2_{\mbox{\fontsize{8}{8}\selectfont av}}
 \label{rozptIMP 1}
\end{equation}
and the right hand side of \eqref{rozptIMP 1} can be easy evaluated:
\begin{eqnarray}
|\langle n' ,k' | U | n, k \rangle|^2_{\mbox{\fontsize{8}{8}\selectfont av}} =
\frac{\gamma^2}{L^2} N_I \left[ \frac{1}{N_I} \sum_{i=1}^{N_I} \chi^2_{n'}(y_i)\chi^2_n(y_i) \right]_{\mbox{\fontsize{8}{8}\selectfont av}}
\nonumber \\ = \frac{\gamma^2 n_I}{L W} \left[ 1 - \frac{\delta_{n n'}}{2} \right], \nonumber \\
\label{scatimp}
\end{eqnarray}
where $n_I = N_I/(WL)$ is the sheet impurity density. To obtain the right hand side of \eqref{scatimp}, we have replaced the term
$\frac{1}{N_I} \sum_{i=1}^{N_I} \chi^2_{n'}(y_i)\chi^2_n(y_i)$ by integral $\frac{1}{W} \int^W_0 \chi^2_{n'}(y)\chi^2_{n}(y) dy$.
Setting \eqref{scatimp} into \eqref{KmatkaA 1}  we get the expression \eqref{Kimp}.

\section{Appendix C: Golden rule for scattering by edge roughness}

Assume first that the electrons are confined in the wire by the potential barriers of finite hight. Specifically, if the wire edges are smooth, the electron Hamiltonian reads
\begin{equation}
H_0 = - \frac{\hbar^2}{2m} \left( \frac{\partial^2}{\partial x^2} + \frac{\partial^2}{\partial y^2} \right) + V_{-}\Theta(-y) + V_{+}\Theta(y-W),
 \label{Hsr}
\end{equation}
where  $V_{-}$ and $V_{+}$ is the hight of the potential barrier at $y=0$ and $y=W$, respectively, and $\Theta$ is the Heaviside step function.
If the wire edges are rough, the Hamiltonian reads $H = H_0 + U$, where
\begin{eqnarray}
U &=& V_{-}\left[ \Theta(-d(x)-y) - \Theta(-y) \right] + \nonumber \\
&+& V_{+}\left[ \Theta(y - h(x)) - \Theta(y-W) \right]
\label{Hsrpo}
\end{eqnarray}
is the perturbation potential due to the edge roughness, with $d(x)$ and $h(x)-W$ being the fluctuations of the edges (see figure \ref{Fig:1}). Now we also assume that $|d(x)| \ll W$ and $|h(x)-W| \ll W$ and we approximate \eqref{Hsrpo} as
\begin{equation}
U = -V_{-}d(x)\delta(-y) - V_{+}[h(x)-W] \delta(y - W).
\label{Hsrpoap}
\end{equation}
Then
\begin{eqnarray}
&|&\hspace{-0.3cm}\langle n' ,k' | U | n, k\rangle|^2_{\mbox{\fontsize{8}{8}\selectfont av}} = \sum_{\beta =\pm} \frac{V^2_{\beta}}{L^2}  \nonumber \\
&\times& \hspace{-0.2cm} \Bigg| \int^L_0 dx \int^W_0 dy e^{i(k - k')x} \chi_{n'}(y) \chi_n(y) d_{\beta}(x) \delta_{\beta}(y)   \Bigg|^2_{\mbox{\fontsize{8}{8}\selectfont av}} \hspace{-0.2cm}, \nonumber \\
\label{rozptERi}
\end{eqnarray}
where $d_{+}(x) \equiv h(x)-W$, $d_{-}(x) \equiv d(x)$, $\delta_+(y) \equiv \delta(y - W)$, and $\delta_-(y) \equiv \delta(y)$. If we integrate in the equation \eqref{rozptERi} over the variable $y$ and then use the limit
\begin{equation}
\lim_{V_{-} \rightarrow \infty} V_{-}\chi^2_n(0) = \lim_{V_{+} \rightarrow \infty} V_{+}\chi^2_n(W) = \frac{\hbar^2 \pi^2 n^2}{m W^3} \equiv A_n ,
\label{vlim}
\end{equation}
both terms in the sum on the right hand side of  \eqref{rozptERi} can be rewritten into the form
\begin{equation}
\frac{A_nA_{n'}}{L^2} \int^L_0 dx_1 \int^L_0 dx_2 e^{i(k - k')(x_1 - x_2)} \langle d_{\beta}(x_1) d_{\beta}(x_2) \rangle.
\label{rozptER}
\end{equation}
where the brackets $\langle \dots \rangle$ symbolize the ensemble averaging (instead of the symbol \emph{av}) and $\langle d_{\beta}(x_1) d_{\beta}(x_2) \rangle$ is the roughness-correlation function. The correlation function depends only on the distance $|x_1 - x_2|$, so we rewrite it as
\begin{equation}
\langle d_{\beta}(x_1) d_{\beta}(x_2) \rangle = \delta^2_{\beta} F_{\beta}(|x_1 - x_2|),
\label{korfu}
\end{equation}
where $F_{\beta}(|x_1 - x_2|)$ is the normalized correlation function  and $\delta_{+}$ and $\delta_{-}$ are the root-mean squares of the randomly fluctuating functions $d_+(x)$ and $d_-(x)$, respectively.
We introduce the variable $x_3 = x_1 - x_2$  and simplify \eqref{rozptER} as
\begin{eqnarray}
&{}& \hspace{-0.6cm} \frac{A_nA_{n'}}{L^2} \delta^2_{\beta} \int^L_0 dx_2 \int^{L-x_2}_{-x_2} dx_3 e^{i(k - k')x_3} F_{\beta}(|x_3|)  \nonumber \\
&\simeq& \frac{A_{n'}A_n}{L} \delta^2_{\beta} \int^{\infty}_{-\infty} dx_3 e^{i(k - k')x_3} F_{\beta}(|x_3|).
\label{rozptERf}
\end{eqnarray}
To obtain the second line in \eqref{rozptERf}, we have replaced the limits $L-x_2$ and $-x_2$ in the first line by $\infty$ and $-\infty$, respectively.
Except for very small $x_2$, such replacement is justified because  $F_{\beta}(|x_3|)$ decays with increasing $x_3$ to zero on the distance (correlation length) much smaller than $L$.

The equation \eqref{rozptERi} can now be expressed as
\begin{equation}
|\langle n' ,k' | U | n, k\rangle|^2_{\mbox{\fontsize{8}{8}\selectfont av}} = \sum_{\beta =\pm} \frac{A_nA_{n'}}{L}
\delta^2_{\beta} \mathcal{F}_{\beta}(|k-k'|),
\label{rozptERfi}
\end{equation}
where
\begin{equation}
\mathcal{F}_{\beta}(q) = \int^{\infty}_{-\infty} e^{iqx_3} F_{\beta}(x_3) dx_3
\label{fourik}
\end{equation}
is the Fourier transform of $F_{\beta}(x_3)$.
Assuming the same randomness at both edges we can put
\begin{equation}
\delta_{\pm} = \delta, \quad F_{\pm}(x_3) = F(x_3), \quad \mathcal{F}_{\pm}(q) = \mathcal{F}(q),
\label{equivedge}
\end{equation}
i.e., we skip the indices $\pm$. We set \eqref{equivedge} into \eqref{rozptERfi} and we replace the symbol $\sum_{\beta =\pm}$ in \eqref{rozptERfi} by the factor of 2.

Finally, we set \eqref{rozptERfi} into \eqref{KmatkaA 1}. Performing summation over $k$ and $k'$ we obtain the matrix elements $K_{n n'}$
in the form \eqref{Ker}.

%


\begin{thebibliography}{99}
\bibitem{comment4} On the first glance it might seem that the scattering has to be weak if $L$ is sufficiently small.
However, the quasi-ballistic (not ballistic) regime is a meaningfull concept only if the length $L$
involves at least one edge step. Even one edge step can scatter strongly if the step size is large.
\bibitem{Mosko} All this is correct for weak impurities. If the impurities are not weak, the semiclassical Boltzmann result
has to be modified by treating the single-impurity scattering non-perturbatively. See e.g. M. Mo\v sko and P. Vagner, Phys. Rev. B \textbf{59}, R10445 (1999).
\end{thebibliography}
\end{document}